\documentclass{article}
\usepackage{amsfonts,amsmath,amssymb,authblk,floatrow,fullpage,graphicx,tabularx,threeparttable,tikz,pgfplots,xcolor,lscape,multirow,verbatim, algorithm, algpseudocode}
\usepackage[colorlinks]{hyperref}
\usepackage[noabbrev,capitalise,sort&compress]{cleveref}
\usepackage{caption,subcaption}
\usepackage{appendix}
\usepackage[textsize = small]{todonotes}
\usepackage[nohyperlinks,nolist]{acronym}
\usepackage{import}
\usetikzlibrary{arrows,arrows.meta,calc,fit,spy}

\renewcommand{\vec}[1]{\boldsymbol{#1}}         
\newcommand{\mat}[1]{\boldsymbol{#1}}           
\newcommand{\pd}{\partial}                      
\newcommand{\parder}[2]{\frac{\pd #1}{\pd #2}} 
\definecolor{ccao}{rgb}{0.596, 0.251, 0.043}
\definecolor{ccaa}{rgb}{0.314, 0.882, 0.957}
\definecolor{ael}{rgb}{0.063, 0.714, 0.188}
\definecolor{elel}{rgb}{0.478, 0.953. 0.373}
\definecolor{gb}{rgb}{0.584, 0.106, 0.478}
\definecolor{cel}{rgb}{0.427, 1.0, 0.420}
\definecolor{cccc}{rgb}{0.796, 0.145, 0.129}
\definecolor{elccc}{rgb}{0.216, 0.114, 0.781}
\definecolor{ccco}{rgb}{0.533, 0.322, 0.055}
\definecolor{color_pot}{rgb}{0.412, 0.976, 1.0}
\definecolor{color_flux}{rgb}{0.412, 0.976, 1.0}
\definecolor{color_curr}{rgb}{0.675, 0, 0.082}
\newfloatcommand{capbtabbox}{table}[][\FBwidth]
\pgfplotsset{compat=newest}
\begin{document}
\title{A conservative and efficient model for grain boundaries of solid electrolytes in a continuum model for solid-state batteries}
\author{Stephan Sinzig$^{1,2}$, Christoph P. Schmidt$^{1}$, Wolfgang A. Wall$^{1}$}
\affil{\small $^1$ Technical University of Munich, Germany, TUM School of Engineering and Design, Institute for Computational Mechanics, Boltzmannstra\ss e 15, 85748 Garching bei M\"unchen \\ $^2$ TUMint.Energy Research GmbH, Lichtenbergstra\ss e 4, 85748 Garching bei M\"unchen, Germany}

\date{}
\maketitle
\section*{Abstract}
A formulation is presented to efficiently model ionic conduction inside, i.e. across and along, grain boundaries. Efficiency and accuracy is achieved by reducing it to a two-dimensional manifold while guaranteeing the conservation of mass and charge at the intersection of multiple grain boundaries. The formulation treats the electric field and the electric current as independent solution variables. We elaborate on the numerical challenges this formulation implies and compare the computed solution with results from an analytical solution by quantifying the convergence towards the exact solution. Towards the end of this work, the model is firstly applied to setups with extreme values of crucial parameters of grain boundaries to study the influence of the ionic conduction in the grain boundary on the overall battery cell voltage and, secondly, to a realistic microstructure to show the capabilities of the formulation.
\section{Introduction}
The technology of solid-state batteries (SSBs) has made remarkable progress in recent years towards their usage in real-world applications. However, some key challenges still remain and are part of current research~\cite{Janek2023}. One of the challenges is to tune the grain boundaries inside polycrystalline solid electrolytes to obtain desired properties. They occur in polycrystalline solid electrolytes between the grains of oxides (e.g., garnet LLZO) or sulfides (e.g., LPS) \cite{Milan2022}. Grains can be defined as the geometric domains where atoms are periodically arranged \cite{Milan2022}, and therefore, grain boundaries are the locations where this periodicity ends. \\
The role of grain boundaries in terms of their influence on the overall cell performance is part of the discussion in the scientific community. Among others, the following physical effects are observed at grain boundaries: The deposition and growth of lithium filaments~\cite{Cheng2017a}, a different ionic conductivity inside of the grain boundary compared to the grains~\cite{Dawson2017,Milan2022}, the grain boundary as an electron conductor in the solid electrolyte~\cite{Song2019} and in the SEI~\cite{Feng2021}, the grain boundary as an ionic resistor~\cite{Dawson2017}, or cracking along the grain boundary~\cite{Milan2022}. In the past years, results have been reported to modify and thereby improve the properties of grain boundaries~\cite{Basappa2017,Zheng2021,Sun2023,Mori2023}. A model that resolves the ionic conduction along the grain boundaries is needed to best profit from the possibility of tuning the properties of grain boundaries. Such a model is also demanded in the literature~\cite{Milan2022}. Such a model allows quantifying the competing transport mechanisms through the grains and through the grain boundaries~\cite{Dawson2017}. Especially for sulfides with high grain boundary conductivity, which could reach the magnitude of the bulk conductivity~\cite{Seino2014}, such a model becomes inevitable. Different types of models have been reported to capture the effect of grain boundaries, ranging from atomistic models \cite{Dawson2017, Dawson2019, Yu2017} to continuum models which discretize the grain boundaries by a phase field \cite{Daubner2023}. However, a continuum model that geometrically resolves the grain boundaries sharply, i.e., not by a smeared phase field, and is applicable to realistic microstructures, is still missing. One challenge is that the thickness of the grain boundary, which is reported to be less than 10 nm \cite{Tenhaeff2013, Rangasamy2012}, is significantly smaller than the typical length scale of a realistic microstructure of a battery cell, and therefore, a standard numerical discretization would exceed the limits of available computational resources (c.f. \cite{Sinzig2023a}). Another challenge comes with the need to incorporate the intersection of multiple grain boundaries at one point while maintaining conservation properties. \\
In this paper, we propose a continuum model that resolves the ionic conduction property of grain boundaries across and along the grain boundary together with a mass and charge-conserving formulation at the intersection of grain boundaries (\cref{sec:model}). Afterwards, the model is analyzed in terms of convergence, and conservation properties before it is applied to realistic microstructures (\cref{sec:examples}). The model is parametrized using experimental studies that differentiate the contribution of measured ionic conductivities into the bulk and the grain boundary \cite{Braun2017,Breuer2015}. \\
Many sources indicate that the ionic conductivity of the grain boundary is usually smaller than the conductivity of the solid electrolyte \cite{Dawson2017,Milan2022,Breuer2015}. However, it was shown that the electrochemical properties of grain boundaries can be tuned using surface modification by coating the grains with another solid electrolyte \cite{Yamada2015} or modifications in the lattice structure \cite{Lee2023}. Thus, we varied the ionic conductivity of the grain boundary in a wide range to quantify its influence on the effective conductivity of the solid electrolyte.
\section{An efficient and conservative model for grain boundaries within SSBs}
\label{sec:model}
In this paper, we present a novel model to include the effect of ionic conduction across and along the grain boundaries into a microstructure-resolved continuum model for solid-state batteries.\\
The effect of grain boundaries on the entire cell performance in terms of ion conduction can only be studied if the grain boundaries are geometrically resolved within the solid electrolyte to also account for potentially preferred conduction paths along the grain boundaries. Therefore, a set of equations describing the conservation of mass and charge in the electrodes, the electrolyte, and the current collectors, as well as inside the grain boundaries, is required.
\subsection{Geometric definitions}
\begin{figure}[ht]
    \centering
    \includegraphics[width = 0.5\textwidth]{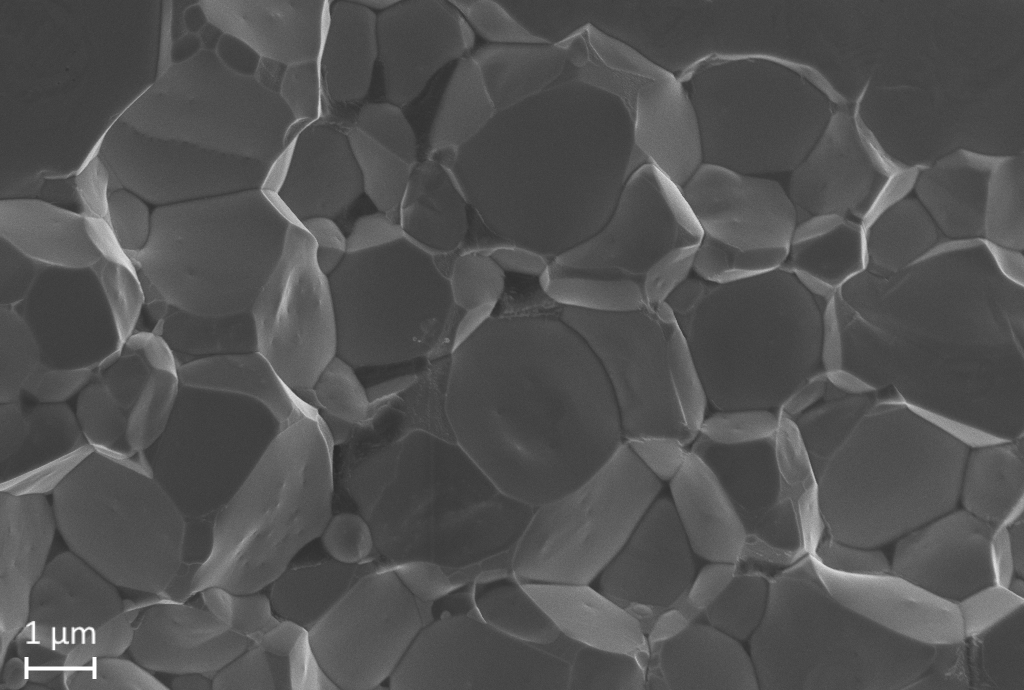}
    \caption{SEM image of the cross-section of LLZO including grains and grain boundaries (M. Balaish, Department of Chemistry, Technical University of Munich).}
    \label{fig:SEM_grain_boundaries}
\end{figure}

In \cref{fig:SEM_grain_boundaries}, an SEM image of grains (bright domains) and grain boundaries (dark lines) of LLZO is shown.
One geometric characteristic of grain boundaries is that more than two can intersect at one line (in the two-dimensional picture at one point). An abstract geometric schematic containing all relevant domains, interfaces, and boundaries of a full battery cell (c.f. \cref{fig:geometry}) includes:
\begin{figure}[ht]
    \centering
    \def\svgwidth{\textwidth}
    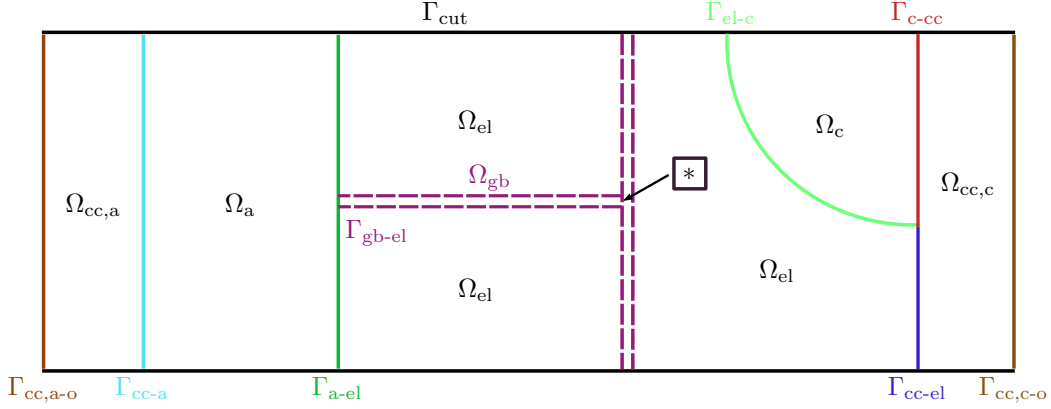
    \caption{Schematic sketch of the geometry of a battery cell including all domains~$\Omega_i$, boundaries~$\Gamma_i$, and interfaces between domains~$\Gamma_{i-j}$.}
    \label{fig:geometry}
  \end{figure}
The current collector on the anode side~$\Omega_\text{cc,a}$, the anode~$\Omega_\text{a}$, the solid electrolyte~$\Omega_\text{el}$, the cathode~$\Omega_\text{c}$, and the current collector on the cathode side~$\Omega_\text{cc,c}$. At the intersection of two domains ($\Omega_i$, $\Omega_j$) interfaces are defined as $\Gamma_{i-j}$. The outer boundary is split into real physical boundaries at the current collectors~$\Gamma_\text{cc-o}$ and modeling boundaries~$\Gamma_\text{cut}$ that are required to limit the size of the geometry in the lateral direction. Additionally, grain boundaries are defined within the solid electrolyte~$\Omega_\text{gb}$ with the interface between them and the solid electrolyte~$\Gamma_\text{gb-el}$. They intersect at one location marked with '*'. Note that this location is a point in two-dimensional setups and lines and points in three-dimensional setups. The electrodes $\Omega_\text{ed} = \Omega_\text{a} \cup \Omega_\text{c}$ and the current collectors $\Omega_\text{cc} = \Omega_\text{cc,a} \cup \Omega_\text{cc,c}$ are summarized to simplify the notation.
\subsection{Governing equations in a continuum formulation}
\begin{figure}[ht]
    \centering
    \def\svgwidth{0.5\textwidth}
    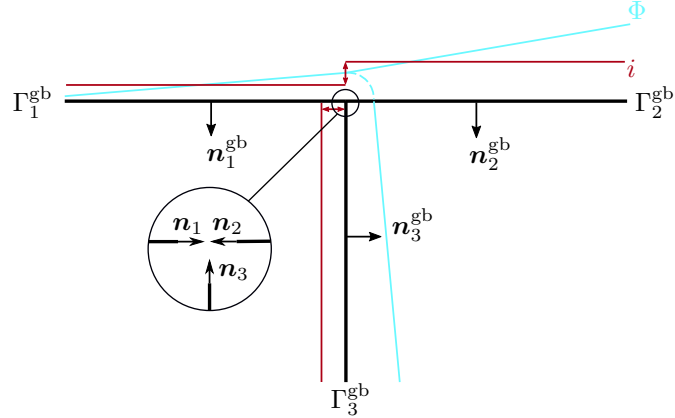
    \caption{Schematic of the intersection of three grain boundaries (black lines) with the definition of the normal vectors and exemplary slopes of the electric potential (blue line) and the electric current (red line), including the constraints at the intersection. The arc illustrates the equal electric potential at this location, and the red arrows sum up to zero, indicating the constraint on the electric current.}
    \label{fig:triple_point}
  \end{figure}

We introduced the set of conservation equations of charge and mass for SSBs in a continuum form in our previous work~\cite{Schmidt2023}. The governing equations are
\begin{subequations}
\begin{align}
    \parder{c}{t} - \nabla \cdot (D \nabla c) = 0 \quad & \text{in} \ \Omega_\text{ed}, \\
    \nabla \cdot (- \sigma \nabla \Phi_\text{ed,cc}) = 0 \quad & \text{in} \ \Omega_\text{ed} \cup \Omega_\text{cc}, \label{eq:laplace_electrode} \\
    \nabla \cdot (- \kappa \nabla \Phi_\text{el}) = 0 \quad & \text{in} \ \Omega_\text{el}, \label{eq:laplace_solid_electrolyte}\\
    \Phi = 0 \quad & \text{on} \ \Gamma_\text{cc,a-o}, \\
    - \vec{i} \cdot \vec{n} = \hat{i} \quad & \text{on} \ \Gamma_\text{cc,c-o}, \\
    \vec{i} \cdot \vec{n} = \vec{j} \cdot \vec{n} = 0 \quad & \text{on} \ \Gamma_\text{cut}, \\
    i = \vec{i}_\text{ed} \cdot \vec{n}_\text{ed} = - \vec{i}_\text{el} \cdot \vec{n}_\text{el} =  i_0 \left[\text{exp} \left(\frac{\alpha_\text{a} F \eta}{R T}\right) - \text{exp} \left(\frac{-(1-\alpha_\text{a}) F \eta}{R T}\right) \right] \quad & \text{on} \ \Gamma_\text{el-ed} \\
    \vec{j}_\text{ed} \cdot \vec{n}_\text{ed} = - \vec{j}_\text{el} \cdot \vec{n}_\text{el} = \frac{F z}{t_+} i \quad & \text{on} \ \Gamma_\text{el-ed} \\
    \vec{i}_\text{ed} \cdot \vec{n}_\text{ed} = - \vec{i}_\text{cc} \cdot \vec{n}_\text{cc}  = \frac{\Phi_\text{cc} - \Phi_\text{ed}}{r_\text{i}} \quad & \text{on} \ \Gamma_\text{ed-cc},\\
    c(t=0) = 
    \begin{cases}
        c_{0,\text{a}} \quad & \text{in} \ \Omega_\text{a} \\
        c_{0,\text{c}} \quad & \text{in} \ \Omega_\text{c}
    \end{cases} &,
\end{align}
\end{subequations}
with the overvoltage $\eta = \Phi_\text{ed} - \Phi_\text{el} - \Phi(c)$. The used symbols are listed in \cref{table:list_of_symbols}.
\begin{table}[ht]
    \begin{tabular}{c | l}
        \textbf{Symbol} & \textbf{Description} \\
        \hline
        $c$ & concentration of species in the electrode \\
        $D$ & diffusion coefficient in the electrode \\
        $F$ & Faraday constant \\
        $\vec{i}$ & electric current density \\
        $i_0$ & exchange current density \\
        $\vec{j}$ & mass flux density \\
        $\vec{n}$ & normal vector \\
        $R$ & universal gas constant \\
        $r_\text{i}$ & interface resistance \\
        $T$ & temperature \\
        $t$ & time \\
        $t_+$ & transference number in the solid electrolyte\\
        $z$ & charge number \\
        $\alpha_a$ & anodic symmetry coefficient \\
        $\kappa$ & ionic conductivity in the solid electrolyte \\ 
        $\Phi$ & electric potential \\
        $\sigma$ & electronic conductivity in the electrodes and current collectors \\       
        \hline
    \end{tabular}
    \caption{List of symbols.}
    \label{table:list_of_symbols}
\end{table} \noindent
Note that we consider the ionic conductivity~$\kappa_\text{el}$ as constant and isotropic inside the grains of the solid electrolyte within this work. However, including an anisotropic and inhomogeneous ionic conductivity would be conceptually easy.\\
This set of governing equations is extended by equations to model the transport of ions in the grain boundaries and the exchange of ions between the grains and the grain boundaries. We model the grain boundaries as single-ion conductors, such that the charge density~$\rho_\text{gb}$ and mass density, expressed in terms of the concentration~$c_\text{gb}$ inside the grain boundaries, are linked by the charge number $z$, the transference number $t_+ = 1$, and the Faraday constant $F$: $\rho_\text{gb} = z F c_\text{gb}$. Therefore, it is sufficient to only model the conservation of charge inside of the grain boundaries. For the derivation of the conservation of charge inside of the grain boundaries, we begin with the general equation for the conservation of charge $\rho_\text{gb}$
\begin{equation}
    \label{eq:balance}
    \parder{\rho_\text{gb}}{t} = \nabla \cdot \vec{i}_\text{gb} + s_{\rho_\text{gb}} \quad \text{in} \ \Omega_\text{gb},
\end{equation}
with the electric current~$\vec{i}_\text{gb}$. The constitutive equation for the electric current is given by
\begin{equation}
    \label{eq:constitutive_flux}
    \vec{i}_\text{gb} = -\kappa_\text{gb} \nabla \Phi_\text{gb} \quad \text{in} \ \Omega_\text{gb},
\end{equation}
to model ion conduction inside of the grain boundary \cite{Milan2022}. No accumulation of charge occurs due to the assumption of electro-neutrality and a transference number of one, i.e.~$\parder{\rho_\text{gb}}{t}=0$.\\
In our previous work \cite{Sinzig2023a}, we showed that thin layers in SSBs can be modeled as two-dimensional manifolds if the transport phenomena that are orthogonal to the layer can be approximated by a priori knowledge. Often, this is given if both the local curvature of the layer and the thickness of the layer are significantly smaller than a typical length scale of the surrounding geometry. A similar concept is used in this work to model the conservation of charge inside of the grain boundaries. Here, we assume that the normal component of the electric field $E_\text{n} = \vec{E} \cdot \vec{n}^\text{gb} = \nabla \Phi \cdot \vec{n}^\text{gb}$ is constant throughout the thickness of the grain boundary while it may vary along the grain boundary (see \cref{fig:triple_point} for the definition of~$\vec{n}^\text{gb}$). This leads to the following set of equations with the Nabla operator $\nabla_\Gamma$ being evaluated on curved surfaces
\begin{subequations}
\begin{alignat}{2}
   -\nabla_\Gamma \cdot \vec{i}_\text{gb} + s_{\rho_\text{gb}} &= 0 &&\quad \text{on} \ \Gamma_\text{gb} \times t_\text{gb}, \label{eq:balance_manifold}\\
   -\kappa_\text{gb} \nabla_\Gamma \Phi_\text{gb} &= \vec{i}_\text{gb} &&\quad \text{on} \ \Gamma_\text{gb} \times t_\text{gb} \label{eq:constitutive_flux_manifold}, \\
   -\vec{i}_\text{gb} \cdot \vec{n} &= \bar{i} &&\quad \text{on} \ \partial {\Gamma_\text{gb}}_\text{h} \times t_\text{gb}, \label{eq:neumann_manifold} \\
   s_\rho &= \frac{\eta}{t_\text{coat} r_\text{n}} + s_{\rho_{\text{gb},0}} &&\quad \text{on} \ \Gamma_\text{gb} \times t_\text{gb},
\end{alignat}
\end{subequations}
with the Neumann boundary of a grain boundary $\partial {\Gamma_\text{gb}}_\text{h}$. The surface $\Gamma_\text{gb}$ is defined in the center of $\Omega_\text{gb}$. From geometric considerations, it becomes obvious that no external fluxes across the outer edges of the grain boundary occur, i.e., $\bar{i} = 0$ on $\Gamma_\text{gb} \cap  \Gamma_\text{cut}$ and $\Gamma_\text{gb} \cap  \Gamma_\text{ed}$. A model for the electric current between the grain boundaries and the solid electrolyte is formulated by a simple linear kinetics law
\begin{equation}
    \vec{i}_\text{el} \cdot \vec{n} = s_{\rho_{\text{gb},0}} = \frac{\Phi_\text{gb} - \Phi_\text{el}}{t_\text{gb} r_\text{c}} \quad \text{on} \ \Gamma_\text{gb} \times t_\text{gb}, \label{eq:coupling_manifold_bulk}
\end{equation}
with the contact resistance~$r_\text{c}$ between the solid electrolyte and the grain boundary domain. However, any other kinetics law is applicable as well. A geometric characteristic of grain boundaries is that more than two can intersect in one line (see the location of the $\ast$ symbol in \cref{fig:geometry}). The conservation of charge needs to be guaranteed at these locations as well. When modeling the grain boundaries as a two-dimensional continuum, the formulation of conservation of charge at these locations is equivalent to Kirchhoff's circuit laws in electrical circuits: (1) the electric potential at the tip of all branches of the grain boundaries is equal, and (2) all currents into and out of the branches have to sum up to zero. This results in the following two constraints wherever~$n$ grain boundaries intersect
\begin{subequations}
\begin{align}
    \Phi &= \Phi_i \ \forall \, i \in n, \label{eq:constr_pot}\\
    \sum_i^n \vec{i}_i \cdot \vec{n}_i &= 0 \ \forall \, n, \label{eq:constr_flux}
\end{align}
\end{subequations}
with the normal vector~$\vec{n}_i$ on the tip of the branches of the grain boundaries~${\Gamma_\text{gb}}_i$ as defined in \cref{fig:triple_point}.
An exemplary slope of the electric potential (blue line) and of the electric current (red line) is added to \cref{fig:triple_point} to visualize the constraints at the intersection point. While the electric potential is equal (denoted by the circular dotted line), the currents need to sum up to zero (denoted by the same length of the red arrows). Note that only the contribution of the flux out of the grain boundary is constrained in a three-dimensional setup.
\subsection{Numerical realization of the model}
First, we summarize the numerical discretization of the novel continuum model. Afterwards, the solution strategy applied to the discretized system of equations is discussed.
\subsubsection{Discretization of the model and solution of the nonlinear system}
For the discretization of the novel model in space, we apply the finite element method and, in time, the one-step theta method. The resulting nonlinear, algebraic system is solved by the iterative Newton-Raphson scheme. \\
We propose to include the constraints formulated in \cref{eq:constr_pot,eq:constr_flux} by treating both the electric potential~$\Phi$ and the electric current~$\vec{i}$ as independent solution variables in the domain of the grain boundary~$\Gamma_\text{gb}$. In the remaining domains, the electric potential~$\Phi$ and the concentration~$c$ are considered as solution variables. This results in more unknowns inside the domain of the grain boundary (i.e., the electric potential and the vector-valued electric current per node). However, we believe that this is affordable as the number of nodes in the grain boundary is significantly smaller than the total number of nodes if the grain boundary is resolved as a two-dimensional manifold. \\
The weighted residual in the domain of the grain boundary consists of the sum of the standard residual~$R_\text{gb,std}$, and the residua originating from the constraints of the electric potential~$R_{\text{gb,constr}_\Phi}$ and of the electric current~$R_{\text{gb,constr}_i}$: $R_\text{gb} = R_\text{gb,std} + R_{\text{gb,constr}_\Phi} + R_{\text{gb,constr}_i} = 0$. The standard residual is given by multiplication of the balance equation (\cref{eq:balance_manifold}), the constitutive equation (\cref{eq:constitutive_flux_manifold}), and the homogeneous Neumann boundary (\cref{eq:neumann_manifold}) with an arbitrary test function $w$ and subsequent integration over the entire domain or boundary, respectively
\begin{equation}
    R_\text{gb,std} = -\int_{\Gamma_\text{gb}} w_\Phi^\text{T} \left(\nabla_\Gamma \cdot \vec{i}_\text{gb} + s_{\rho_\text{gb}} \right) \ \text{d} \Gamma + \int_{\Gamma_\text{gb}} \vec{w}_i^\text{T} (\vec{i}_\text{gb} + \kappa_\text{gb} \nabla_\Gamma \Phi_\text{gb}) \ \text{d} \Gamma + \int_{\partial {\Gamma_\text{gb}}_\text{h}} w_\Phi^\text{T} \left( \vec{i}_\text{gb} \cdot \vec{n}\right) \ \text{d} \partial \Gamma.
\end{equation}
Applying Gauss divergence theorem and the product rule of the divergence to the first term leads to
\begin{subequations}
\begin{equation}
    R_\text{gb,std} = \int_{\Gamma_\text{gb}} \nabla_\Gamma w_\Phi^\text{T} \vec{i}_\text{gb} \ \text{d} \Gamma + \int_{\Gamma_\text{gb}} w_\Phi^\text{T} s_{\rho_\text{gb}} \ \text{d} \Gamma + \int_{\Gamma_\text{gb}} \vec{w}_i^\text{T} (\vec{i}_\text{gb} + \kappa_\text{gb} \nabla_\Gamma \Phi_\text{gb}) \ \text{d} \Gamma,
    \label{eq:residual_std}
\end{equation}
where we make use of $w_\Phi = 0$ on $\partial \Gamma_\text{gb} \backslash \partial {\Gamma_\text{gb}}_\text{h}$.
The constraints (\cref{eq:constr_pot,eq:constr_flux}) are enforced by introducing Lagrange multipliers
\begin{align}
    R_{\text{gb,constr}_\Phi} &= \sum_{j=1}^{n-1} \left( \int_{\partial \Gamma_\text{gb}} {w_{{\lambda_\Phi},j}}^\text{T} \left( \Phi_\text{m} - \Phi_{\text{s},j} \right) \ \text{d} \partial \Gamma + \int_{\partial \Gamma_\text{gb}} \left({w_\Phi}_\text{m}^\text{T} - {{w_\Phi}_{\text{s},j}}^\text{T}\right) \lambda_{\Phi,j} \ \text{d} \partial \Gamma \right) \label{eq:residual_constr_phi},\\
    R_{\text{gb,constr}_i} &= \int_{\partial \Gamma_\text{gb}} w_{\lambda_i}^\text{T} \left( i_\text{s}^1 + \mat{C}_i \vec{i}_\text{m} \right) \ \text{d} \partial \Gamma + \int_{\partial \Gamma_\text{gb}} \left( {{w_i}_\text{s}^1}^\text{T} + {\vec{w}_i}_\text{m}^\text{T} \mat{C}_i^\text{T} \right) \lambda_i \ \text{d} \partial \Gamma \label{eq:residual_constr_i}.
\end{align}
\end{subequations}
The constraint matrix~$\mat{C}_i$ is chosen, such that the flux constraint (\cref{eq:constr_flux}) is satisfied. Therefore, the constraint is split into a slave and a master side and reorganized with~${n_\text{s}}^1$ being the largest component of the normal vector~$\vec{n}_\text{s}$ ($n_\text{s}^1 = \text{max}(n_\text{s}^x, n_\text{s}^y, n_\text{s}^z)$), and~${n_\text{s}}^2, {n_\text{s}}^3$ the other components
\begin{equation}
    i_\text{s}^1 + \sum_{j=2}^{n_\text{dim}} i_\text{s}^j \frac{n_\text{s}^j}{n_\text{s}^1} + \sum_{k=1}^{n-1} \sum_{j=1}^{n_\text{dim}} i_{\text{m},k}^j \frac{n_{\text{m},k}^j}{n_\text{s}^1} = i_\text{s}^1 + \mat{C}_i \vec{i}_\text{m} = 0.
    \label{eq:constr_flux_reorg}
\end{equation}
In the remaining domains, only the electric potential and the concentration are considered unknown, leading to the following discretized form of the weighted residual
\begin{equation}
    \begin{split}
    R_\text{bulk} = 
    \kappa_\text{el} \int_{\Omega_\text{el}} \nabla w_\Phi^\text{T} \nabla \Phi_\text{bulk} \ \text{d} \Omega + \int_{{\Gamma_\text{el}}_\text{h}} w_\Phi^\text{T} \bar{i} \ \text{d} \Gamma
    + \sigma \int_{\Omega_\text{ed,cc}} \nabla w_\Phi^\text{T} \nabla \Phi_\text{bulk} \ \text{d} \Omega + \int_{{\Gamma_\text{ed,cc}}_\text{h}} w_\Phi^\text{T} \bar{i} \ \text{d} \Gamma \\
    + \int_{\Omega_\text{ed}} w_c^\text{T} \parder{c}{t} \ \text{d} \Omega + D \int_{\Omega_\text{ed}} \nabla w_c^\text{T} \nabla c \ \text{d} \Omega + \int_{{\Gamma_\text{ed}}_h} w_c^\text{T} \bar{j} \ \text{d} \Gamma = 0. \label{eq:residual_bulk}
    \end{split}
\end{equation}
The residua~$R_\text{gb}$ and~$R_\text{bulk}$ are discretized in space using the finite element method, and the resulting nonlinear, algebraic system is solved using the Newton-Raphson scheme (see Appendix~\ref{sec:remarks_constr} for the derivation of the linear system of equations in the domain of the grain boundaries and Appendix~\ref{sec:remarks_bulk} for the remaining domains). Thereafter, two linearized systems of equations ($\mat{K}_\text{gb} \Delta \vec{\Psi}_\text{gb} = - \vec{R}_\text{gb}$, $\mat{K}_\text{bulk} \Delta \vec{\Psi}_\text{bulk} = - \vec{R}_\text{bulk}$) arise
\begin{subequations}
    \begin{align}
        \mat{K}_\text{gb} &=
        \begin{bmatrix}
            {\mat{K}_\text{gb}}_{\Phi \Phi} & {\mat{K}_\text{gb}}_{\Phi i} \\
            {\mat{K}_\text{gb}}_{i \Phi} & {\mat{K}_\text{gb}}_{i i}
        \end{bmatrix}\\
        \vec{R}_\text{gb} &=
        \begin{bmatrix}
            {\vec{R}_\text{gb}}_\Phi &
            {\vec{R}_\text{gb}}_i
        \end{bmatrix}^\text{T},
    \end{align}
\end{subequations}
\begin{subequations}
\begin{align}
    \vec{R}_\text{bulk} =
    \begin{bmatrix}
        {\vec{R}_\text{bulk}}_\Phi \\
        {\vec{R}_\text{bulk}}_c
    \end{bmatrix}, \\
    \mat{K}_\text{bulk} =
    \begin{bmatrix}
        {\mat{K}_\text{bulk}}_{\Phi \Phi} & {\mat{K}_\text{bulk}}_{\Phi c} \\
        {\mat{K}_\text{bulk}}_{c \Phi} & {\mat{K}_\text{bulk}}_{c c}
    \end{bmatrix},
\end{align}
\end{subequations}
The systems of equations for the grain boundaries and the bulk domains are coupled by the coupling flux defined in \cref{eq:coupling_manifold_bulk}. Therefore, the systems of equations contain the additional contributions $\mat{K_{{\text{bulk}_\Phi}{\text{gb}_\Phi}}}$ and $\mat{K_{{\text{gb}_\Phi}{\text{bulk}_\Phi}}}$
\begin{equation}
    \begin{bmatrix}
        {\mat{K}_\text{bulk}}_{\Phi \Phi} & {\mat{K}_\text{bulk}}_{\Phi c} & \mat{K_{{\text{bulk}_\Phi}{\text{gb}_\Phi}}} & \mat{0} \\
        {\mat{K}_\text{bulk}}_{c \Phi} & {\mat{K}_\text{bulk}}_{c c} & \mat{0} & \mat{0} \\
        \mat{0} & \mat{K_{{\text{gb}_\Phi}{\text{bulk}_\Phi}}} & {\mat{K}_\text{gb}}_{\Phi \Phi} & {\mat{K}_\text{gb}}_{\Phi i} \\
        \mat{0} & \mat{0} & {\mat{K}_\text{gb}}_{i \Phi} & {\mat{K}_\text{gb}}_{i i} \\
    \end{bmatrix}
    \begin{bmatrix}
        \Delta \vec{\hat{c}}_\text{bulk} \\
        \Delta \vec{\hat{\Phi}}_\text{bulk} \\
        \Delta \vec{\hat{\Phi}}_\text{gb} \\
        \Delta \vec{\hat{i}}_\text{gb}
    \end{bmatrix}
    =
    -
    \begin{bmatrix}
        \vec{R}_{c_\text{bulk}} \\
        \vec{R}_{\Phi_\text{bulk}} \\
        \vec{R}_{\Phi_\text{gb}} \\
        \vec{R}_{i_\text{gb}} \\
    \end{bmatrix}.
    \label{eq:assembled_linear_system}
\end{equation}
\subsubsection{Solution of the linear system}
Due to robustness and efficiency, this linear system of equations (\cref{eq:assembled_linear_system}) is solved monolithically~\cite{Verdugo2016}. Iterative solvers are required as soon as the size of the linear system of equations exceeds a certain threshold, which is often the case for realistic microstructures. This monolithic system could include a zero-block on the main-diagonal sub-block~${\mat{K}_\text{gb}}_{\Phi \Phi}$ if the source term~$s_{\rho_\text{gb}}$ is not a function of the electric potential, i.e.~$\parder{s_{\rho_\text{gb}}}{\vec{\hat{\Phi}}_\text{gb}}=0$. Many iterative solvers fail for such saddle-point problems, i.e. systems that contain a zero sub-block on the main diagonal (c.f. \cite{Benzi2005}). However, the determinant of the entire matrix~$\mat{K}$ is non-zero if proper boundary conditions are applied, such that a solution exists. \\
In this work, we adapted an approach that employs the Block-Gauss-Seidel (BGS) algorithm (see \cref{alg:BGS}, adapted from \cite{Fang2019}). Therefore, the linear system of equations (\cref{eq:assembled_linear_system}) is split into physically meaningful sub-blocks, which, in this case, are the bulk domains (i.e., the electrodes, solid electrolyte, and current collectors) and the domain of the grain boundary
\begin{equation}
    \begin{bmatrix}
        \mat{K}_{11} & \dots & \mat{K}_{1n} \\
        \vdots & \ddots & \vdots \\
        \mat{K}_{n1} & \dots & \mat{K}_{nn}
    \end{bmatrix}
    \begin{bmatrix}
        \Delta \vec{x}_1 \\
        \vdots \\
        \Delta \vec{x}_n
    \end{bmatrix}
    =
    \begin{bmatrix}
        \vec{R}_1 \\
        \vdots \\
        \vec{R}_n
    \end{bmatrix}.
\end{equation}
Application of the BGS algorithm results in multiple smaller systems of linear equations that only require inverting the main-diagonal blocks. We know that the main diagonal block of the grain boundary domain could have a saddle-point structure. Therefore, the system of equations of this sub-block is solved with a direct solver (in this case UMFPACK \cite{Davis2004}) while the other sub-blocks are solved using an algebraic multigrid solver (c.f.~\cite{Xu2017a}) together with a row and column-based equilibration. This is affordable, as for realistic microstructures, the number of unknowns in the grain boundary block is significantly smaller (for our example in \cref{sec:realisitc_example}: $6,200 \ \text{nodes} \cdot 4 \frac{\text{dofs}}{\text{node}} = 24,800 \ \text{dofs}$) compared to the number of unknowns in the other domains ($157,000 \ \text{nodes} \cdot 2 \frac{\text{dofs}}{\text{node}} = 314,000 \ \text{dofs}$), and thus, the computationally greater amount of workload of direct solvers does not weigh heavy. \\
Note, that several other methods are available to circumvent the solution of a saddle-point system, e.g., by making use of the Schur-complement (c.f. \cite{Benzi2005} or \cite{Rozloznik2018}). However, the focus of this work is not on the assessment and comparison of different solvers.
\subsection{Remarks on alternative formulations}
Alternative formulations exist to enforce the constraint on the electric current using the finite-element method. One possibility is to change the function space of the shape functions to Hermite shape functions (c.f.~\cite{Wriggers2008} for details). These shape functions include the derivative at the nodal values, which scales with the electric current, as an additional unknown. Thereby, constraints on the electric current could easily be applied. However, this change of the function space requires a fundamental reformulation of finite element codes and is often not applicable. Another possibility is given by enforcing the constraint in standard formulations using Lagrangian shape functions. However, the derivative of a quantity includes, in general, all nodes of an element, and therefore, the constraint affects not just the nodes at the interface $\Gamma_\text{gb-el}$ but all elemental nodes, which adds additional hurdles to the implementation.
\section{Numerical examples}
\label{sec:examples}
We first analyze the novel model and show the correctness of the model and its implementation for a geometrically simplified setup. Afterwards, the influence of ionically conducting grain boundaries on the effective conductivity of solid electrolytes is discussed for a simple geometry. Finally, results for grain boundaries in realistic microstructures are shown.
\subsection{Analysis of the novel model}
The correctness of the model is shown by a convergence study with an analytical solution as a reference and the evaluation of the conservation of charge at the intersection between three grain boundaries.
\subsubsection{Convergence study on simple grain geometries}
\begin{figure}[ht]
    \RawFloats
    \centering
    \captionbox{Geometry for the convergence study. The grain boundaries occur at the intersection of the solid electrolyte domains (colored rectangles). The edge size of each grain is~$a=4$ or~$2a$. The coordinates~$x_i$ are defined along the grain boundaries. The electric potential is set as a boundary condition at the ends of the grain boundaries marked with a circle.
    \label{fig:t-shape_geometry}}
    [0.45\textwidth]{\def\svgwidth{0.4\textwidth}
    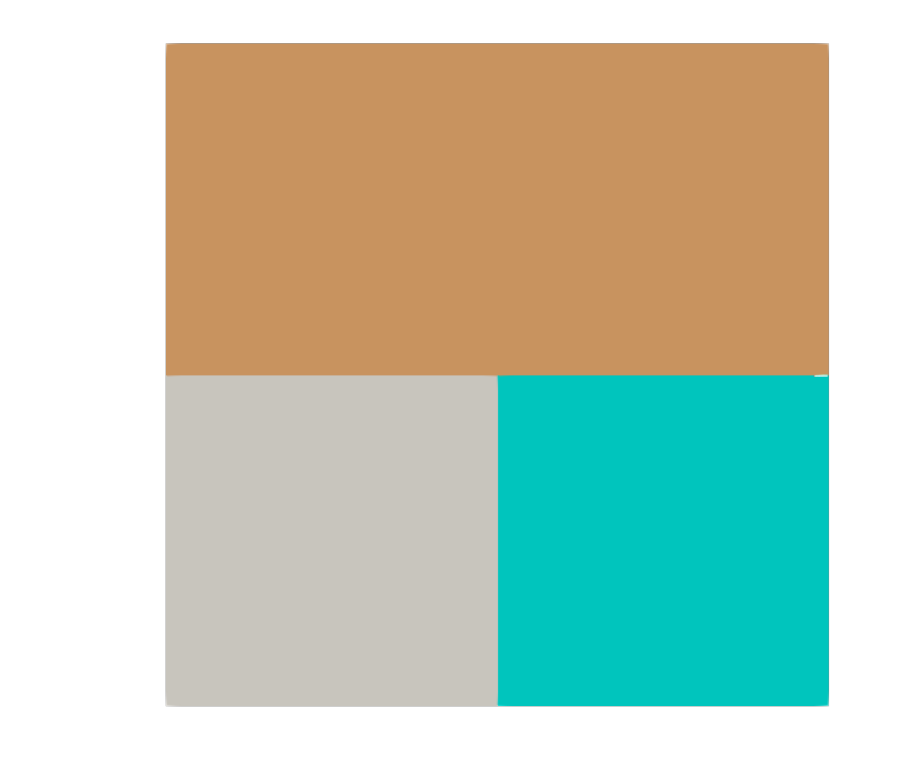}%
    \hfill
    \captionbox{Analytic solution of the electric potential (blue) and the current (orange). The solid line is along the coordinates~$x_1$ and~$x_2$, and the dashed line is along the coordinate~$x_3$ of \cref{fig:t-shape_geometry}. The vectors denote the sign of the normal direction of the three grain boundaries ($\Gamma_1$ - $\Gamma_3$) at the intersection of each of the three grain boundaries. The direction of the flux for positive values is in the positive coordinate direction and vice versa.\label{fig:t-shape_analytic_solution_combined}}
    [0.45\textwidth]{
%
%
\definecolor{mycolor1}{rgb}{0.00000,0.44700,0.74100}%
\definecolor{mycolor2}{rgb}{0.85000,0.32500,0.09800}%
\begin{tikzpicture}
\begin{axis}[%
width=5.0cm,
height=5.0cm,
scale only axis,
xmin=0,
xmax=8,
xtick={0,2,4,6,8},
x tick label style={
/pgf/number format/.cd,
/tikz/.cd,
yshift=-.5em},
xlabel style={font=\color{white!15!black}},
xlabel={position},
every outer y axis line/.append style={mycolor1},
every y tick label/.append style={font=\color{mycolor1}},
every y tick/.append style={mycolor1},
ymin=0,
ymax=4,
ytick={0,1,2,3,4},
y tick label style={
/pgf/number format/.cd,
/tikz/.cd},
ylabel style={font=\color{mycolor1}},
ylabel={electric potential},
axis background/.style={fill=white},
axis x line*=bottom,
axis y line*=left,
xmajorgrids,
ymajorgrids
]
\addplot [color=mycolor1, dashed, line width=1.0pt, forget plot]
  table[row sep=crcr]{%
0	0\\
0.1	0.0138634969758692\\
0.2	0.0277408586040442\\
0.3	0.0416459634026385\\
0.4	0.0555927176352481\\
0.5	0.0695950692183739\\
0.6	0.0836670216705004\\
0.7	0.0978226481167817\\
0.8	0.112076105363339\\
0.9	0.126441648055246\\
1	0.140933642932364\\
1.1	0.155566583197276\\
1.2	0.170355103009699\\
1.3	0.185313992121863\\
1.4	0.200458210669497\\
1.5	0.215802904133209\\
1.6	0.231363418485228\\
1.7	0.24715531553666\\
1.8	0.263194388500591\\
1.9	0.27949667778662\\
2	0.296078487042602\\
2.1	0.312956399459656\\
2.2	0.330147294356739\\
2.3	0.34766836406137\\
2.4	0.365537131103392\\
2.5	0.383771465738961\\
2.6	0.40238960382229\\
2.7	0.421410165043026\\
2.8	0.44085217154749\\
2.9	0.460735066962406\\
3	0.481078735840154\\
3.1	0.501903523544973\\
3.2	0.523230256600025\\
3.3	0.545080263515651\\
3.4	0.567475396119662\\
3.5	0.590438051410987\\
3.6	0.613991193958533\\
3.7	0.638158378867676\\
3.8	0.662963775337324\\
3.9	0.688432190831132\\
4	0.714589095887033\\
};
\addplot [color=mycolor1, line width=1.0pt, forget plot]
  table[row sep=crcr]{%
0	4\\
0.1	3.86744701675554\\
0.2	3.73876180282584\\
0.3	3.61381566227282\\
0.4	3.49248363854343\\
0.5	3.37464438950258\\
0.6	3.26018006608098\\
0.7	3.14897619441618\\
0.8	3.04092156136923\\
0.9	2.93590810330223\\
1	2.83383079800569\\
1.1	2.7345875596676\\
1.2	2.63807913677907\\
1.3	2.54420901287457\\
1.4	2.45288331000744\\
1.5	2.36401069486407\\
1.6	2.27750228742301\\
1.7	2.19327157206757\\
1.8	2.11123431106292\\
1.9	2.03130846031139\\
2	1.95341408730151\\
2.1	1.87747329116887\\
2.2	1.80341012478873\\
2.3	1.73115051882255\\
2.4	1.66062220764256\\
2.5	1.59175465706\\
2.6	1.52447899378515\\
2.7	1.45872793654823\\
2.8	1.39443572881257\\
2.9	1.33153807301258\\
3	1.2699720662508\\
3.1	1.20967613738981\\
3.2	1.15058998547591\\
3.3	1.09265451943319\\
3.4	1.03581179896747\\
3.5	0.980004976621253\\
3.6	0.925178240921458\\
3.7	0.871276760563341\\
3.8	0.818246629574605\\
3.9	0.766034813404936\\
4	0.714589095887033\\
};
\addplot [color=mycolor1, line width=1.0pt, forget plot]
  table[row sep=crcr]{%
4	0.714589095887033\\
4.1	0.690372256395477\\
4.2	0.666845846693256\\
4.3	0.643986338410067\\
4.4	0.621770870132606\\
4.5	0.600177224541243\\
4.6	0.579183806190858\\
4.7	0.558769619913589\\
4.8	0.538914249821922\\
4.9	0.519597838891094\\
5	0.50080106910042\\
5.1	0.482505142113661\\
5.2	0.464691760479117\\
5.3	0.447343109330656\\
5.4	0.430441838571361\\
5.5	0.413971045521987\\
5.6	0.397914258016872\\
5.7	0.3822554179304\\
5.8	0.366978865117539\\
5.9	0.352069321752386\\
6	0.337511877049075\\
6.1	0.323291972349739\\
6.2	0.309395386564649\\
6.3	0.295808221949933\\
6.4	0.282516890208672\\
6.5	0.26950809890148\\
6.6	0.256768838152946\\
6.7	0.244286367640681\\
6.8	0.232048203853933\\
6.9	0.220042107609034\\
7	0.208256071809197\\
7.1	0.196678309436421\\
7.2	0.185297241763486\\
7.3	0.174101486774267\\
7.4	0.163079847780762\\
7.5	0.152221302225479\\
7.6	0.141514990657953\\
7.7	0.130950205874393\\
7.8	0.120516382209589\\
7.9	0.110203084970361\\
8	0.1\\
};
\end{axis}

\begin{axis}[%
width=5.0cm,
height=5.0cm,
scale only axis,
every outer y axis line/.append style={mycolor2},
every y tick label/.append style={font=\color{mycolor2}},
every y tick/.append style={mycolor2},
xmin=0,
xmax=8,
axis x line=none,
ymin=-0.5,
ymax=1.5,
ytick={-0.5, 0, 0.5, 1, 1.5},
y tick label style={
/pgf/number format/.cd,
/tikz/.cd},
ylabel style={font=\color{mycolor2}},
ylabel={current},
axis x line*=bottom,
axis y line*=right
]
\addplot [color=mycolor2, dashed, line width=1.0pt, forget plot]
  table[row sep=crcr]{%
0	-0.138611866625795\\
0.1	-0.138681178334795\\
0.2	-0.138889182779279\\
0.3	-0.139236087981028\\
0.4	-0.139722240874151\\
0.5	-0.140348127652057\\
0.6	-0.141114374253682\\
0.7	-0.142021746989484\\
0.8	-0.143071153307816\\
0.9	-0.144263642702449\\
1	-0.145600407762155\\
1.1	-0.147082785363395\\
1.2	-0.148712258007306\\
1.3	-0.150490455302326\\
1.4	-0.152419155593937\\
1.5	-0.154500287743162\\
1.6	-0.156735933055583\\
1.7	-0.159128327362823\\
1.8	-0.161679863258561\\
1.9	-0.16439309249133\\
2	-0.16727072851647\\
2.1	-0.170315649209819\\
2.2	-0.173530899745822\\
2.3	-0.176919695642961\\
2.4	-0.180485425979543\\
2.5	-0.184231656783057\\
2.6	-0.188162134596505\\
2.7	-0.192280790225249\\
2.8	-0.196591742668152\\
2.9	-0.201099303236914\\
3	-0.205807979867748\\
3.1	-0.210722481629685\\
3.2	-0.215847723434045\\
3.3	-0.221188830949748\\
3.4	-0.226751145729418\\
3.5	-0.232540230551377\\
3.6	-0.238561874982885\\
3.7	-0.244822101170195\\
3.8	-0.251327169861197\\
3.9	-0.258083586666689\\
4	-0.26509810856653\\
};
\addplot [color=mycolor2, line width=1.0pt, forget plot]
  table[row sep=crcr]{%
0	1.34530727007729\\
0.1	1.30597331276865\\
0.2	1.26794543760753\\
0.3	1.23118561354965\\
0.4	1.19565707770755\\
0.5	1.16132429858456\\
0.6	1.12815294054041\\
0.7	1.09610982945268\\
0.8	1.06516291953993\\
0.9	1.03528126131326\\
1	1.00643497062421\\
1.1	0.978595198778167\\
1.2	0.951734103683219\\
1.3	0.925824822005772\\
1.4	0.900841442304972\\
1.5	0.8767589791191\\
1.6	0.853553347978031\\
1.7	0.831201341316756\\
1.8	0.809680605265885\\
1.9	0.788969617295913\\
2	0.769047664692897\\
2.1	0.749894823844015\\
2.2	0.731491940312296\\
2.3	0.713820609680582\\
2.4	0.696863159145583\\
2.5	0.680602629843595\\
2.6	0.665022759890228\\
2.7	0.650107968117161\\
2.8	0.635843338489681\\
2.9	0.622214605189402\\
3	0.609208138347258\\
3.1	0.596810930412498\\
3.2	0.585010583144053\\
3.3	0.573795295211258\\
3.4	0.563153850391544\\
3.5	0.553075606353273\\
3.6	0.543550484012526\\
3.7	0.534568957453174\\
3.8	0.526122044400174\\
3.9	0.518201297236539\\
4	0.510798794555022\\
};
\addplot [color=mycolor2, line width=1.0pt, forget plot]
  table[row sep=crcr]{%
4	0.245700685988491\\
4.1	0.238676464569108\\
4.2	0.231890939504662\\
4.3	0.225337324704611\\
4.4	0.219009066007999\\
4.5	0.21289983462876\\
4.6	0.207003520826394\\
4.7	0.201314227795723\\
4.8	0.195826265769592\\
4.9	0.190534146328631\\
5	0.185432576912372\\
5.1	0.180516455526256\\
5.2	0.175780865639206\\
5.3	0.171221071266689\\
5.4	0.166832512234336\\
5.5	0.162610799617391\\
5.6	0.158551711351415\\
5.7	0.154651188009873\\
5.8	0.150905328744369\\
5.9	0.147310387383474\\
6	0.143862768686236\\
6.1	0.140559024746649\\
6.2	0.13739585154545\\
6.3	0.134370085645833\\
6.4	0.131478701029743\\
6.5	0.128718806071605\\
6.6	0.126087640646464\\
6.7	0.123582573369623\\
6.8	0.121201098965043\\
6.9	0.118940835759856\\
7	0.116799523302496\\
7.1	0.114775020102056\\
7.2	0.112865301486622\\
7.3	0.11106845757843\\
7.4	0.10938269138383\\
7.5	0.107806316996142\\
7.6	0.106337757909609\\
7.7	0.104975545442761\\
7.8	0.103718317269609\\
7.9	0.102564816057208\\
8	0.101513888208218\\
};
\end{axis}
\draw[-Stealth, draw = mycolor2] (2.5,2.53) -- (3.3,2.53);
\draw[-Stealth, draw = mycolor2] (2.5,1.86) -- (1.7,1.86);
\draw[-Stealth, draw = mycolor2, dashed] (2.5,0.6) -- (3.3,0.6);
\node[mycolor2] at (3.3,2.3) {$n_1$};
\node[mycolor2] at (1.9,2.1) {$n_2$};
\node[mycolor2] at (3.2,0.4) {$n_3$};
\end{tikzpicture}%
}
\end{figure}

An analytic solution can be found for certain geometric setups and boundary conditions. This analytic solution serves as a reference for a spatial convergence analysis. Here, a three-dimensional setup without units is chosen where three planar grain boundaries are modeled between solid electrolyte grains (see \cref{fig:t-shape_geometry}). They intersect at one line (depicted in the quasi-two-dimensional setup in \cref{fig:t-shape_geometry} as a point). The coordinates~$x_1$, $x_2$, and~$x_3$ are introduced along the grain boundaries.
The electric potential in the grains is fixed to zero $\Phi_\text{se} = 0$. Thus, the equation for the conservation of charge inside the grain boundaries reduces to a one-dimensional differential equation
\begin{equation}
    \kappa_\text{gb} \frac{\partial^2 \Phi_\text{gb}}{\partial x_i^2} - \frac{\Phi_\text{gb}}{t_\text{gb} \ r_\text{n}} = 0 \quad \forall \ i \in 1,2,3,
\end{equation}
with the second term being the exchange current between the grain boundary and the grains. The analytic solution of the electric potential~$\Phi_\text{gb,ana}$ is given by
\begin{equation}
    \Phi_\text{gb,ana} = c_i^1 \ \text{exp}\left( \frac{x_i}{\sqrt{ \kappa_\text{gb} \ t_\text{gb} \ r_\text{n}}} \right) + c_i^2 \ \text{exp}\left( \frac{-x_i}{\sqrt{ \kappa_\text{gb} \ t_\text{gb} \ r_\text{n}}} \right) \quad \forall \ i \in 1,2,3,
\end{equation}
with the constants~$c_i^1, c_i^2$. Consequently, the current~$i_\text{gb,ana}$ is
\begin{equation}
    i_\text{gb,ana} = - \kappa_\text{gb} \parder{\Phi_\text{gb}}{x_i} = - \kappa_\text{gb} \left[ \frac{c_i^1}{\sqrt{ \kappa_\text{gb} \ t_\text{gb} \ r_\text{n}}} \text{exp}\left( \frac{x_i}{\sqrt{ \kappa_\text{gb} \ t_\text{gb} \ r_\text{n}}} \right) - \frac{c_i^2}{\sqrt{\kappa_\text{gb} \ t_\text{gb} \ r_\text{n}}} \text{exp}\left( - \frac{x_i}{\sqrt{ \kappa_\text{gb} \ t_\text{gb} \ r_\text{n}}} \right) \right].
\end{equation}
By enforcing the constraints defined in \cref{eq:constr_pot,eq:constr_flux} and applying Dirichlet boundary conditions for the electric potential at all ends of the T-shape ($\Phi_1 = 0, \ \Phi_2 = 0.1, \ \Phi_3 = 4$, see circles in \cref{fig:t-shape_geometry}), the constants~$c_i^1, c_i^2$ can be computed. The analytic solution of the electric potential and the electric current within the three grain boundaries~$\Gamma_1$ - $\Gamma_3$ are plotted in \cref{fig:t-shape_analytic_solution_combined} for~$\kappa_\text{gb} = 1$, and $\sqrt{\kappa_\text{gb} t_\text{gb} r_\text{n}} = \sqrt{10}$. The positive direction of the current is defined in the positive direction of the respective coordinate. As expected, the currents sum up to zero, i.e., $i_1 n_1 + i_2 n_2 + i_3 n_3 = 0$, with~$i_i$ denoting the current of each grain boundary~$\Gamma_i$ at the intersection. Note that the direction of the normal vectors of the three grain boundary domains w.r.t. the coordinate has to be considered when computing the sum. A spatial convergence study is performed based on this analytic solution. Therefore, the relative L2-norm~$\epsilon$ of the deviation of the numerical solution of the electric potential~$\Phi_\text{gb}$ from the analytic solution is computed as
\begin{equation}
    \epsilon = \frac{\sqrt{\int \left(\Phi_\text{gb,ana} - \Phi_\text{gb} \right)^2 \ \text{d} x}}{\sqrt{\int \Phi_\text{gb,ana}^2 \ \text{d} x}}.
\end{equation}
The development of the relative L2-norm for a decreasing edge length of the hexahedral elements used to mesh the three-dimensional geometry is shown in \cref{fig:t-shape_convergence_study} together with two lines representing linear and quadratic convergence, respectively.
\begin{figure}[H]
    \centering
%
%
\definecolor{mycolor1}{rgb}{0.00000,0.44700,0.74100}%
\definecolor{mycolor2}{rgb}{0.85000,0.32500,0.09800}%
\definecolor{mycolor3}{rgb}{0.4940, 0.1840, 0.5560}%

\begin{tikzpicture}

\begin{axis}[%
width=7.5cm,
height=5.0cm,
scale only axis,
xmode=log,
xmin=5.0e-2,
xmax=1.0,
xminorticks=true,
xlabel style={font=\color{white!15!black}},
xlabel={Edge length of elements $\Delta l$},
x tick label style={
/pgf/number format/.cd,
/tikz/.cd,
yshift=-.5em},
ymode=log,
ymin=1.0e-5,
ymax=1.0e-1,
y tick label style={
/pgf/number format/.cd,
/tikz/.cd},
yminorticks=true,
ylabel style={font=\color{white!15!black}},
ylabel={Relative L2-norm $\epsilon$},
axis background/.style={fill=white},
xmajorgrids,
xminorgrids,
ymajorgrids,
yminorgrids
]
\addplot [color=mycolor1, line width=1.0pt, mark size=4.0pt, mark=x, mark options={solid, mycolor1}, forget plot]
  table[row sep=crcr]{%
1	0.0154746793751826\\
0.5	0.00386212966427903\\
0.25	0.000965273448852231\\
0.125	0.000241300618075189\\
0.0625	6.03273761017302e-05\\
};
\addplot [color=mycolor2, dashed, line width=1.0pt, forget plot]
  table[row sep=crcr]{%
1	0.012291974751013\\
0.5	0.00307299368775325\\
0.25	0.000768248421938314\\
0.125	0.000192062105484578\\
0.0625	4.80155263711446e-05\\
};
\addplot [color=mycolor3, dashed, line width=1.0pt, forget plot]
  table[row sep=crcr]{%
1	0.0194814671047844\\
0.5	0.00974073355239221\\
0.25	0.0048703667761961\\
0.125	0.00243518338809805\\
0.0625	0.00121759169404903\\
};
\end{axis}
\node[mycolor2] at (3.5,1.65) {$\propto \Delta l^2$};
\node[mycolor3] at (3.5,3.6) {$\propto \Delta l$};
\end{tikzpicture}%
    \caption{Convergence of the relative L2-norm (blue line). Model evaluations are marked with a cross. The dashed lines represent linear and quadratic convergence rates, respectively.}
    \label{fig:t-shape_convergence_study}
\end{figure}
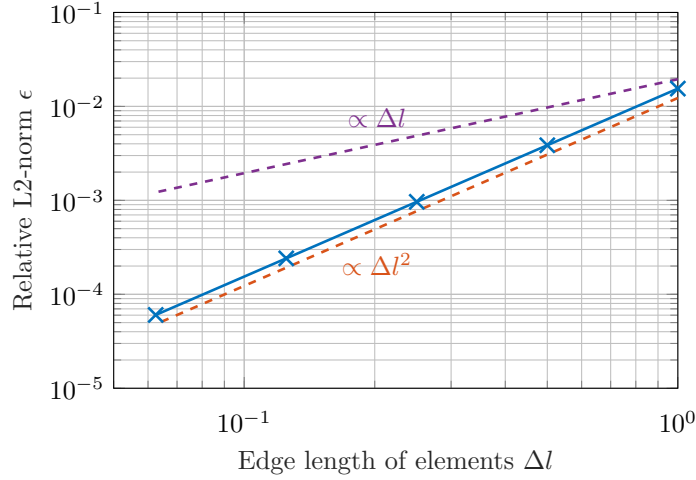 \noindent
As expected, quadratic convergence is observed, which confirms a correct formulation and implementation of the constraints. 
\subsubsection{Conservation at non-orthogonally intersecting grain boundaries}
\begin{figure}[ht]
    \RawFloats
    \centering
    \captionbox{Geometry with non-orthogonally intersecting grain boundaries. The grain boundaries occur at the intersection of the solid electrolyte domains (colored domains with a radius of $r=1$). The electric potential is set as a boundary condition at the ends of the grain boundaries marked with circles.
    \label{fig:cylinders_geometry}}
    [0.45\textwidth]{\def\svgwidth{0.35\textwidth}
\begingroup%
  \makeatletter%
  \providecommand\color[2][]{%
    \errmessage{(Inkscape) Color is used for the text in Inkscape, but the package 'color.sty' is not loaded}%
    \renewcommand\color[2][]{}%
  }%
  \providecommand\transparent[1]{%
    \errmessage{(Inkscape) Transparency is used (non-zero) for the text in Inkscape, but the package 'transparent.sty' is not loaded}%
    \renewcommand\transparent[1]{}%
  }%
  \providecommand\rotatebox[2]{#2}%
  \newcommand*\fsize{\dimexpr\f@size pt\relax}%
  \newcommand*\lineheight[1]{\fontsize{\fsize}{#1\fsize}\selectfont}%
  \ifx\svgwidth\undefined%
    \setlength{\unitlength}{357.07554374bp}%
    \ifx\svgscale\undefined%
      \relax%
    \else%
      \setlength{\unitlength}{\unitlength * \real{\svgscale}}%
    \fi%
  \else%
    \setlength{\unitlength}{\svgwidth}%
  \fi%
  \global\let\svgwidth\undefined%
  \global\let\svgscale\undefined%
  \makeatother%
  \begin{picture}(1,0.96703769)%
    \lineheight{1}%
    \setlength\tabcolsep{0pt}%
    \put(0,0){\includegraphics[width=\unitlength,page=1]{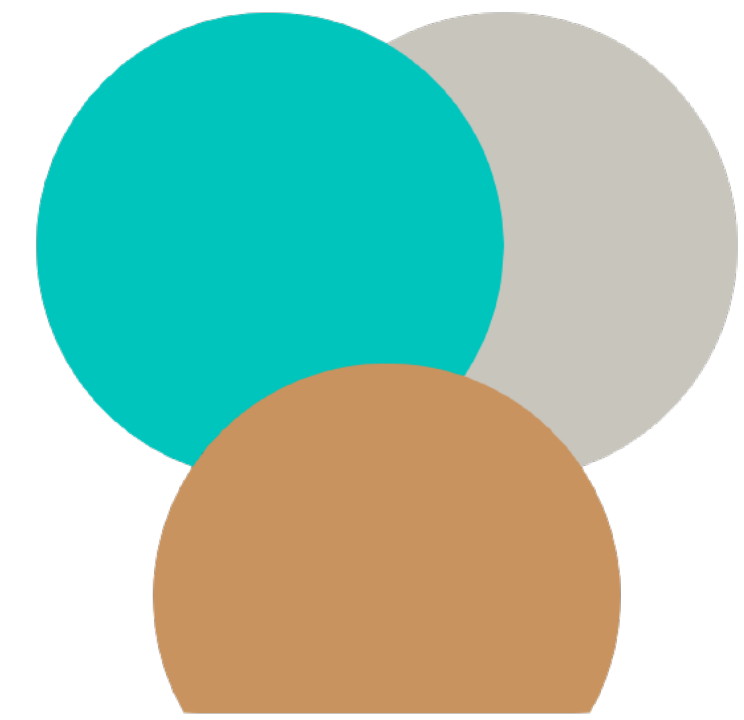}}%
    \put(-0.00185973,0.31385933){\color[rgb]{0,0,0}\makebox(0,0)[lt]{\lineheight{1.25}\smash{\begin{tabular}[t]{l}$\Phi_2=0.1$\end{tabular}}}}%
    \put(0.82405946,0.31329582){\color[rgb]{0,0,0}\makebox(0,0)[lt]{\lineheight{1.25}\smash{\begin{tabular}[t]{l}$\Phi_3=2$\end{tabular}}}}%
    \put(0.44658823,0.95001573){\color[rgb]{0,0,0}\makebox(0,0)[lt]{\lineheight{1.25}\smash{\begin{tabular}[t]{l}$\Phi_1=0$\end{tabular}}}}%
    \put(0,0){\includegraphics[width=\unitlength,page=2]{cylinders_geometry.pdf}}%
  \end{picture}%
\endgroup%
}%
    \hfill
    \captionbox{Current in the grain boundary. The jump at the intersection, s.t. the net current sums up to zero, is visible. Note that the geometry is distorted for visualization compared to \cref{fig:cylinders_geometry}. \label{fig:cylinders_flux_in_gb}}
    [0.45\textwidth]{\includegraphics[width = 0.35\textwidth]{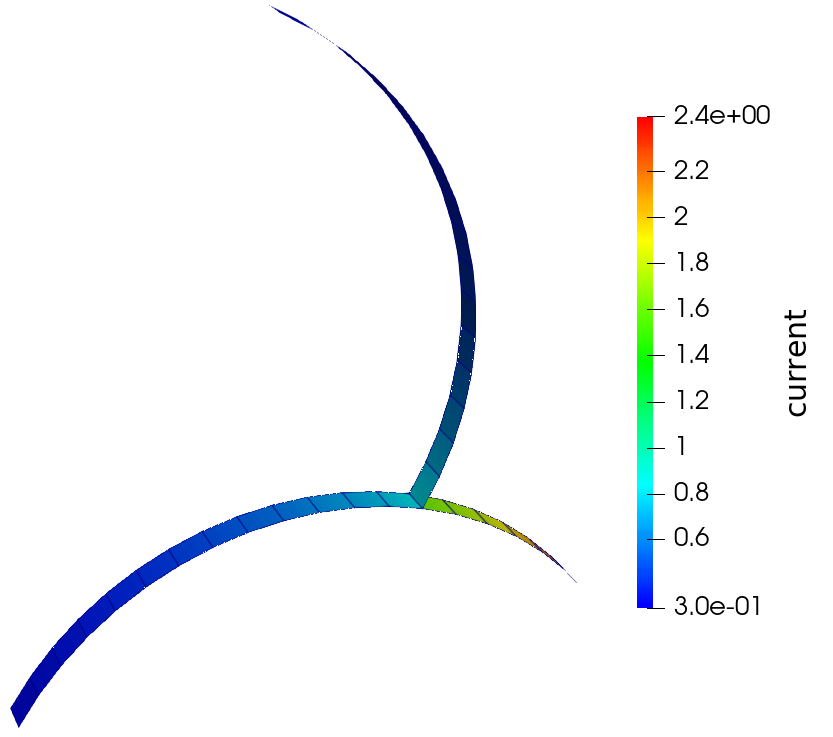}}
\end{figure}

A geometry including non-orthogonal intersections of the grain boundaries is investigated in another three-dimensional example, again without units. The solid electrolyte grains are modeled as three intersecting cylindrical grains (see circles in \cref{fig:cylinders_geometry}). 
The electric potential in the solid electrolyte grains is fixed to zero $\Phi_\text{se} = 0$. At the outer ends of the grain boundaries, the electric potential is fixed ($\Phi_1 = 0, \ \Phi_2 = 0.1, \ \Phi_3 = 2$, see \cref{fig:cylinders_geometry}) to different values and the ionic conductivity of the grain boundaries is set to $\kappa=0.1$. For this setup, the sum of the currents at the intersection of the grain boundaries can be evaluated to quantify the fulfillment of the constraints. Therefore, the relative sum of the electric currents into the grain boundaries~$\epsilon = \frac{i_\text{sum}}{i_\text{tot}}$ is computed at the intersection of the grain boundaries with~$i_\text{sum} = \sum_i \vec{n}_i \cdot \vec{i}_i$, and the total current~$i_\text{tot} = \sum_i |\vec{i}_i|$. For the outlined case, the electric current is shown in \cref{fig:cylinders_flux_in_gb} and the relative sum is~$\epsilon = \frac{i_\text{sum}}{i_\text{tot}}= 6.8 \cdot 10^{-9}$ at the intersection, which shows the fulfillment of the charge conservation at the intersection within the expected numerical tolerances.
\subsection{Assessment of the influence of the grain boundaries}
\begin{figure}[ht]
    \centering
    \def\svgwidth{0.8\textwidth}
\begingroup%
  \makeatletter%
  \providecommand\color[2][]{%
    \errmessage{(Inkscape) Color is used for the text in Inkscape, but the package 'color.sty' is not loaded}%
    \renewcommand\color[2][]{}%
  }%
  \providecommand\transparent[1]{%
    \errmessage{(Inkscape) Transparency is used (non-zero) for the text in Inkscape, but the package 'transparent.sty' is not loaded}%
    \renewcommand\transparent[1]{}%
  }%
  \providecommand\rotatebox[2]{#2}%
  \newcommand*\fsize{\dimexpr\f@size pt\relax}%
  \newcommand*\lineheight[1]{\fontsize{\fsize}{#1\fsize}\selectfont}%
  \ifx\svgwidth\undefined%
    \setlength{\unitlength}{267.07162302bp}%
    \ifx\svgscale\undefined%
      \relax%
    \else%
      \setlength{\unitlength}{\unitlength * \real{\svgscale}}%
    \fi%
  \else%
    \setlength{\unitlength}{\svgwidth}%
  \fi%
  \global\let\svgwidth\undefined%
  \global\let\svgscale\undefined%
  \makeatother%
  \begin{picture}(1,0.22625921)%
    \lineheight{1}%
    \setlength\tabcolsep{0pt}%
    \put(0,0){\includegraphics[width=\unitlength,page=1]{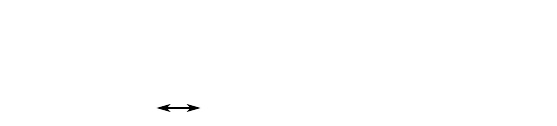}}%
    \put(0.30969473,0.00706447){\color[rgb]{0,0,0}\makebox(0,0)[lt]{\lineheight{1.25}\smash{\begin{tabular}[t]{l}$t_\text{el}$\end{tabular}}}}%
    \put(0,0){\includegraphics[width=\unitlength,page=2]{regular_grains_geometry.pdf}}%
    \put(0.15825526,0.2035008){\color[rgb]{0,0,0}\makebox(0,0)[lt]{\lineheight{1.25}\smash{\begin{tabular}[t]{l}$z$\end{tabular}}}}%
    \put(0,0){\includegraphics[width=\unitlength,page=3]{regular_grains_geometry.pdf}}%
    \put(1.00248646,0.09739597){\color[rgb]{0,0,0}\makebox(0,0)[lt]{\lineheight{1.25}\smash{\begin{tabular}[t]{l}$\Phi_2$\end{tabular}}}}%
    \put(0,0){\includegraphics[width=\unitlength,page=4]{regular_grains_geometry.pdf}}%
    \put(-0.00248646,0.09761688){\color[rgb]{0,0,0}\makebox(0,0)[lt]{\lineheight{1.25}\smash{\begin{tabular}[t]{l}$\Phi_1$\end{tabular}}}}%
    \put(0,0){\includegraphics[width=\unitlength,page=5]{regular_grains_geometry.pdf}}%
  \end{picture}%
\endgroup%

    \caption{Regularly arranged solid electrolyte grains (colored domains). The grain boundaries occur between the solid electrolyte grains. The difference in electric potential is $\Delta \Phi_\text{SE} = \Phi_2 - \Phi_1$.}
    \label{fig:regular_grains_geometry}
\end{figure}

\begin{figure}[ht]
    \centering
    \begin{subfigure}[b]{0.45\textwidth}
        \centering
%
%
\definecolor{mycolor1}{rgb}{0.0,0.0,0.0}%
\definecolor{mycolor3}{rgb}{0.25,0.25,0.25}%
\definecolor{mycolor4}{rgb}{0.5,0.5,0.5}%
\definecolor{mycolor5}{rgb}{0.46600,0.67400,0.18800}%
\begin{tikzpicture}

\begin{axis}[%
width=5.0cm,
height=5.0cm,
scale only axis,
xmode=log,
xmin=1e-08,
xmax=1000,
xtick={1e-8, 1e-6, 1e-4, 1e-2, 1e0, 1e2},
xminorticks=true,
xlabel style={font=\color{white!15!black}},
xlabel={ionic conductivity in $\frac{\text{S}}{\text{m}}$},
x tick label style={
/pgf/number format/.cd,
/tikz/.cd,
yshift=-.5em},
ymode=log,
ymin=0.05,
ymax=5.0,
yminorticks=true,
ylabel style={font=\color{white!15!black}},
ylabel={diff. in el. pot. in solid electrolyte in V},
y tick label style={
/pgf/number format/.cd,
/tikz/.cd},
axis background/.style={fill=white},
xmajorgrids,
ymajorgrids,
yminorgrids
]
\addplot [color=mycolor1, dashed, line width=1.0pt, forget plot]
  table[row sep=crcr]{%
1e-07	3.19272576812226\\
1.25892541179417e-07	2.72382697822386\\
1.58489319246111e-07	2.3513674301795\\
1.99526231496888e-07	2.05551229487523\\
2.51188643150958e-07	1.82050620751487\\
3.16227766016838e-07	1.63383423699246\\
3.98107170553497e-07	1.48555542017489\\
5.01187233627272e-07	1.36777336936518\\
6.30957344480193e-07	1.2742157608633\\
7.94328234724281e-07	1.19990031085698\\
1e-06	1.14086945064071\\
1.25892541179417e-06	1.09397957165087\\
1.58489319246111e-06	1.05673361684644\\
1.99526231496888e-06	1.02714810331601\\
2.51188643150958e-06	1.00364749457998\\
3.16227766016838e-06	0.984980297527735\\
3.98107170553497e-06	0.970152415845978\\
5.01187233627272e-06	0.958374210765007\\
6.30957344480193e-06	0.949018449914819\\
7.94328234724282e-06	0.941586904914187\\
1e-05	0.93568381889256\\
1.25892541179417e-05	0.930994830993576\\
1.58489319246111e-05	0.927270235513133\\
1.99526231496888e-05	0.92431168416009\\
2.51188643150958e-05	0.921961623286486\\
3.16227766016838e-05	0.920094903581262\\
3.98107170553497e-05	0.918612115413086\\
5.01187233627272e-05	0.917434294904989\\
6.30957344480194e-05	0.916498718819971\\
7.94328234724282e-05	0.915755564319907\\
0.0001	0.915165255717745\\
0.000125892541179417	0.914696356927846\\
0.000158489319246111	0.914323897379802\\
0.000199526231496888	0.914028042244498\\
0.000251188643150958	0.913793036157137\\
0.000316227766016838	0.913606364186615\\
0.000398107170553497	0.913458085369797\\
0.000501187233627272	0.913340303318988\\
0.000630957344480194	0.913246745710486\\
0.000794328234724282	0.913172430260479\\
0.001	0.913113399400263\\
0.00125892541179417	0.913066509521273\\
0.00158489319246111	0.913029263566469\\
0.00199526231496888	0.912999678052938\\
0.00251188643150958	0.912976177444202\\
0.00316227766016838	0.91295751024715\\
0.00398107170553497	0.912942682365468\\
0.00501187233627272	0.912930904160387\\
0.00630957344480193	0.912921548399537\\
0.00794328234724281	0.912914116854537\\
0.01	0.912908213768515\\
0.0125892541179417	0.912903524780616\\
0.0158489319246111	0.912899800185136\\
0.0199526231496888	0.912896841633782\\
0.0251188643150958	0.912894491572909\\
0.0316227766016838	0.912892624853204\\
0.0398107170553497	0.912891142065036\\
0.0501187233627272	0.912889964244528\\
0.0630957344480193	0.912889028668442\\
0.0794328234724281	0.912888285513942\\
0.1	0.91288769520534\\
0.125892541179417	0.91288722630655\\
0.158489319246111	0.912886853847002\\
0.199526231496888	0.912886557991867\\
0.251188643150958	0.91288632298578\\
0.316227766016838	0.912886136313809\\
0.398107170553497	0.912885988034992\\
0.501187233627272	0.912885870252941\\
0.630957344480193	0.912885776695333\\
0.794328234724281	0.912885702379883\\
1	0.912885643349023\\
1.25892541179417	0.912885596459144\\
1.58489319246111	0.912885559213189\\
1.99526231496888	0.912885529627675\\
2.51188643150958	0.912885506127067\\
3.16227766016838	0.91288548745987\\
3.98107170553497	0.912885472631988\\
5.01187233627272	0.912885460853783\\
6.30957344480193	0.912885451498022\\
7.94328234724281	0.912885444066477\\
10	0.912885438163391\\
12.5892541179417	0.912885433474403\\
15.8489319246111	0.912885429749808\\
19.9526231496888	0.912885426791256\\
25.1188643150958	0.912885424441195\\
31.6227766016838	0.912885422574476\\
39.8107170553497	0.912885421091687\\
50.1187233627273	0.912885419913867\\
63.0957344480193	0.912885418978291\\
79.4328234724282	0.912885418235136\\
100	0.912885417644828\\
};
\addplot [color=mycolor1, dashed, line width=1.0pt, forget plot]
  table[row sep=crcr]{%
1e-07	0.083061497931823\\
100	0.083061497931823\\
};
\addplot [color=mycolor3, line width=1.0pt, forget plot]
  table[row sep=crcr]{%
0.0786	0.06\\
0.0786	4.0\\
};
\addplot [color=mycolor4, line width=1.0pt, forget plot]
  table[row sep=crcr]{%
0.0188	0.06\\
0.0188	4.0\\
};
\addplot [color=mycolor5, line width=1.0pt, mark size=6.0pt, mark=asterisk, mark options={solid, mycolor5}, forget plot]
  table[row sep=crcr]{%
1e-07	3.1365113943069\\
1e-06	1.12052053320522\\
1e-05	0.916951390509501\\
0.0001	0.884277731133695\\
0.001	0.838631391401089\\
0.01	0.714722689510686\\
0.1	0.377284704105202\\
1	0.129387476407176\\
10	0.0865519780264224\\
100	0.0819510937478621\\
};
\end{axis}
\draw[-Stealth] (3.8,0.3) -- (4.7,0.3);
\draw[-Stealth] (1.1,3.8) -- (0.7,4.7);
\end{tikzpicture}%
        \caption{The extreme cases (infinite conductivity and zero conductivity in the grain boundaries) are shown by dashed lines, and evaluations of the proposed model are represented by the green line. The ionic conductivity of the grains (dark grey) and the grain boundary (light grey) are indicated by vertical lines for LLTO~\cite{Braun2017}.}
        \label{fig:regular_grains_pot_diff}
    \end{subfigure}
    \hfill
    \begin{subfigure}[b]{0.45\textwidth}
        \centering
        \input{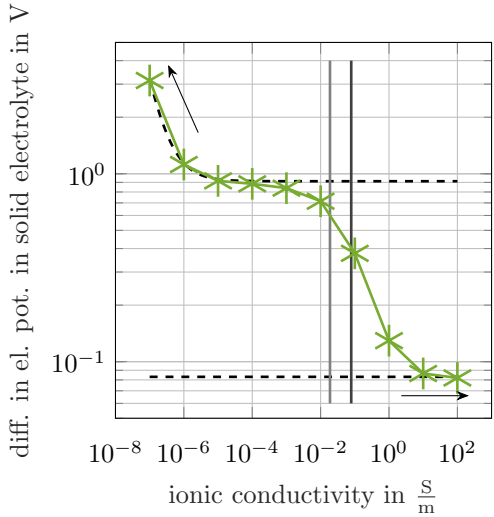}
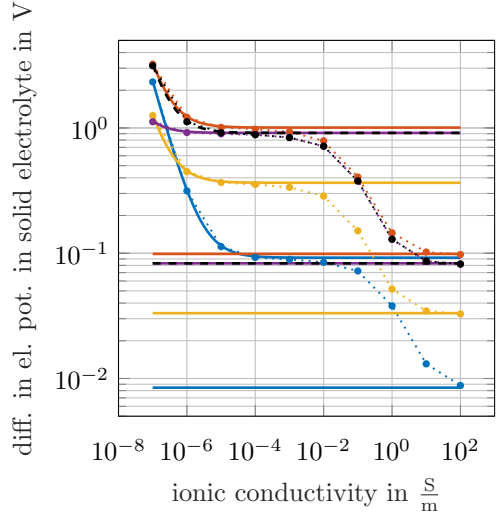
        \caption{Variation of parameters that influence the extreme cases. The dashed black line represents the default values and the colored lines the variations of the parameters in the following manner: blue: $r_\text{c} < {r_\text{c}}_0$, red: $\kappa_\text{se} < {\kappa_\text{se}}_0$, yellow: $i < i_0$, purple: $t_\text{gb} < {t_\text{gb}}_0$, where the index 0 represents the default value. The evaluations with the novel model are symbolized by the dotted lines.}
        \label{fig:regular_grains_pot_case_study}
    \end{subfigure}
    \caption{Evaluation of the influence of the ionic conductivity of the grain boundary on the voltage drop in the solid electrolyte.}
\end{figure}

An artificial geometry with regularly arranged grains, as shown in \cref{fig:regular_grains_geometry}, allows finding analytical results for extreme cases of the ionic conductivity inside of the grain boundary.
These results serve as a basis to gain deeper insights and enable comparison with the solution obtained from the simulation. We analyze the difference in electric potential~$\Delta \Phi_\text{SE}$ through the solid electrolyte, i.e., from the leftmost point of the solid electrolyte to the rightmost point, for a given current~$i$ through the solid electrolyte. The difference in electric potential is a measure of the effective conductivity of the solid electrolyte. We distinguish between two extreme cases: (1) the ionic conductivity inside the grain boundaries approaches zero, and (2) the ionic conductivity inside the grain boundaries approaches infinity. In the first extreme case, conduction inside the grain boundaries is unfavored, and therefore, the shortest conduction path is exclusively in the z-direction, i.e., it is strictly orthogonal across the grain boundaries, which are normal to the z-direction. For this case, the difference in electric potential can be estimated by a series of resistors of $n_\text{SE}=12$ solid electrolyte grains ($r_\text{se}=n_\text{se} \frac{t_\text{se}}{\kappa_\text{se}}$), $n_\text{gb}=11$ grain boundaries ($r_\text{gb}=n_\text{gb} \frac{t_\text{gb}}{\kappa_\text{gb}}$), and~$2 n_\text{gb}$ contact resistances ($r_\text{gb-se}=2 \ n_\text{gb} r_\text{c}$) scaled by the current~$i$
\begin{subequations}
\begin{equation}
    \Delta \Phi_\text{se,zero} = i \left(n_\text{SE} \frac{t_\text{SE}}{\kappa_\text{SE}} + n_\text{gb} \left[\frac{t_\text{gb}}{\kappa_\text{c}} + 2 \ r_\text{c} \right] \right).
    \label{eq:upper_limit}
\end{equation}
In the second extreme case, conduction in the grains is unfavored. Therefore, the shortest conduction path is through the grain boundaries because their resistance approaches zero in the limit of infinite ionic conductivity. Thus, the remaining resistance originates from two solid electrolyte grains ($r_\text{se} = 2 \frac{t_\text{se}}{\kappa_\text{se}}$) and two contact resistances ($r_\text{gb-se} = 2 r_\text{c}$)
\begin{equation}
    \Delta \Phi_\text{se,inf} = i \left(2 \frac{t_\text{se}}{\kappa_\text{se}}  + 2 \ r_\text{c} \right).
    \label{eq:lower_limit}
\end{equation}
\end{subequations}
The solution of both equations is plotted in \cref{fig:regular_grains_pot_diff} in dashed lines for different values of the ionic conductivity in the grain boundary~$\kappa_\text{gb}$.
The other parameters are constant as defined in \cref{table:parameters_regular_grains}.
\begin{table}[ht]
    \begin{tabular}{c | l}
        \textbf{Symbol} & \textbf{Description} \\
        \hline
        $\kappa_\text{gb}$ & $[10^{-7}, \ 10^2] \frac{\text{S}}{\text{m}}$ \\
        $\kappa_\text{se}$ & $7.86 \cdot 10^{-2} \frac{\text{S}}{\text{m}}$ \\
        $t_\text{se}$ & $3 \ \mu\text{m}$ \\
        $t_\text{gb}$ & $ 10 \ \text{nm}$ \\
        $r_\text{c}$ & $2 \cdot 10^{-2} \ \Omega \text{m}^2$ \\        
        $i$ & $2.07 \ \frac{\text{A}}{\text{m}^2}$  \\
        \hline
    \end{tabular}
    \caption{Parameters for the setup with the regular grains.}
    \label{table:parameters_regular_grains}
\end{table}
It can be seen that both curves are separated from each other, and especially if the focus of the investigation lies between the extreme cases, a model that resolves conduction along the grain boundary becomes vital. Vertical lines exemplarily indicate the values for the conductivity of LLTO inside the grains (dark grey) and inside the grain boundary (light grey). They highlight that the incorporation of conduction inside grain boundaries is important for realistic material parameters. The difference in electric potential of evaluations of the novel model with different values for the ionic conductivity of the grain boundary~$\kappa_\text{gb}$ are added to \cref{fig:regular_grains_pot_diff}. It can be seen that the proposed model is able to cover both extreme cases with a smooth transition between them. This transition is within a relevant range of the ionic conductivity of typical solid electrolyte materials and grain boundaries and, therefore, requires resolving charge transport also along the grain boundary. Based on these two extreme cases, a deeper understanding of the ionic conduction in the grain boundary can be derived. Therefore, we modify single parameters from \cref{eq:upper_limit,eq:lower_limit} and plot the results in \cref{fig:regular_grains_pot_case_study} together with the results of evaluations with the novel model. It can be seen that the characteristic shape remains the same for all combinations while the magnitude and slope change:
\begin{itemize}
    \item Reduction of the contact resistance (blue curve, reduction by one order of magnitude) reduces the total resistance. This is relevant within a large range of the ionic conductivity of the grain boundaries. For small ionic conductivities of the grain boundaries, the total resistance is dominated by this small ionic conductivity, and therefore, the influence of the contact resistance becomes negligible in these regions.
    \item Reduction of the ionic conductivity in the solid electrolyte (red curve, reduction by two orders of magnitude) increases the total resistance. However, this is only relevant for high ionic conductivities of the grain boundary as for small ionic conductivities, the total resistance is dominated by this small ionic conductivity.
    \item Reduction of the current (yellow curve, reduction by factor 2.5) parallelly shifts the difference in electric potential. For this setup, the current is simply a linear amplification.
    \item Reduction of the thickness of the grain boundary (purple curve, reduction by one order of magnitude) reduces the total resistance for small ionic conductivities of the grain boundary while it remains the same for high ionic conductivities.
\end{itemize}
We can conclude that the novel model is capable of resolving all variations, including the extreme cases.
\subsection{Application of the model to a realisitc microstructure}
\label{sec:realisitc_example}
The focus of this work is to establish a formulation to include ion conduction in intersecting grain boundaries into a continuum model for SSBs. Hence, for completeness, we show its applicability to a realistic and, therefore, complex microstructure.
\subsubsection{Microstructure}
\begin{figure}[ht]
    \centering
    \begin{subfigure}[b]{0.4\textwidth}
        \centering
        \def\svgwidth{\textwidth}
        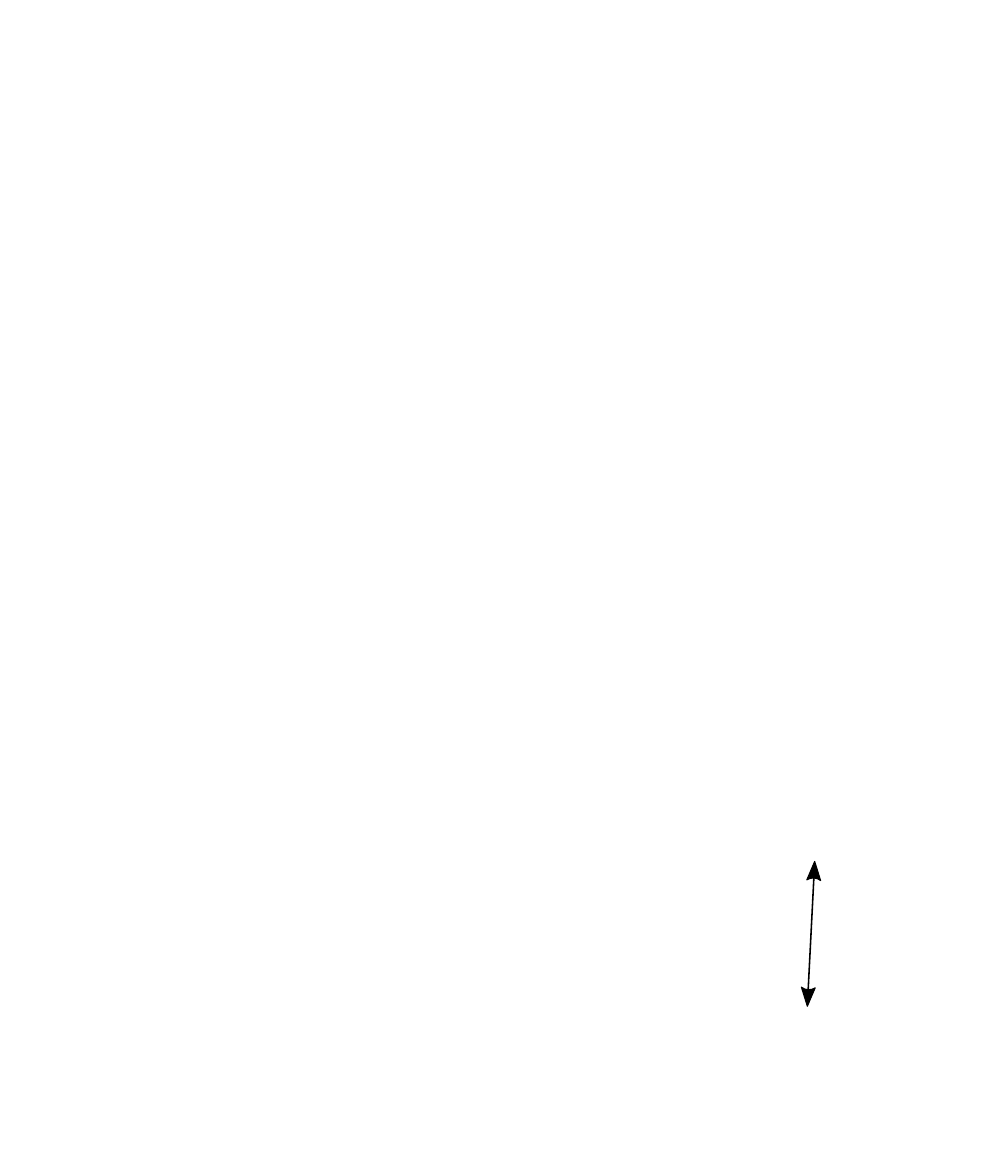
        \caption{Battery cell with porous separator and composite electrode.}
        \label{fig:realistic_geometry_full}
    \end{subfigure}
    \hfill
    \begin{subfigure}[b]{0.4\textwidth}
        \centering
        \includegraphics[width=\textwidth]{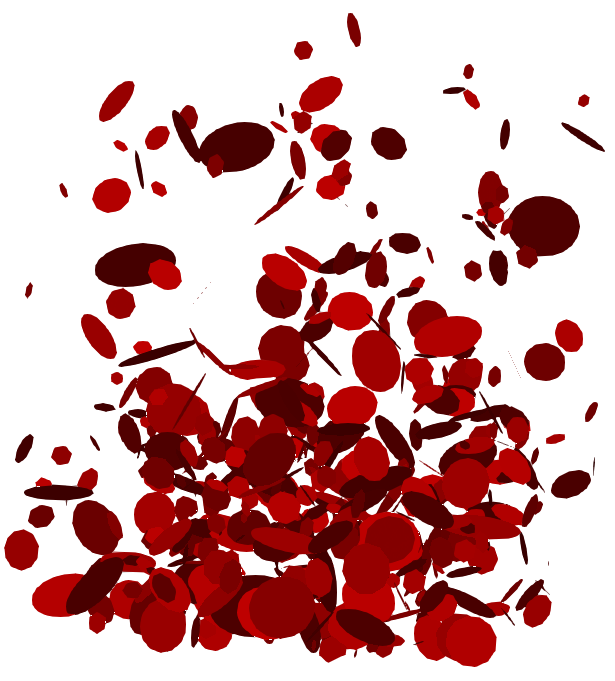}
        \caption{Network of grain boundaries.}
        \label{fig:realistic_geometry_gb}
    \end{subfigure}
    \caption{Geometry of battery cell with grain boundaries. The grain boundaries occur at the intersections of the solid electrolyte grains and are modeled as planar surfaces.}
\end{figure}

The grains are approximated as spherical particles (see \cref{fig:realistic_geometry_full}, blue: solid electrolyte, grey: cathode active material, red: grain boundaries, silver: anode, silver/brown: current collectors).
The composite cathode consists of active material particles (mean of diameter distribution $\mu_\text{c} = 2.3$, and standard deviation $\sigma_\text{c} = 0.05$~\cite{Neumann2020}, evaluated in $\mu$m) and solid electrolyte grains ($\mu_\text{el} = 1.831$, $\sigma_\text{el} = 0.548$~\cite{Fu2023}, evaluated in $\mu$m) which follow a log-normal distribution for the diameter~$d$ with the probability density function
\begin{equation}
    p = \frac{1}{d \sigma \sqrt{2 \pi}} \text{exp} \left( \frac{- \left(\text{ln}(d) - \mu \right)^2}{2 \sigma^2} \right)
\end{equation}
Both phases have a volumetric ratio of $\frac{v_\text{c}}{v_\text{c} + v_\text{el}} = 0.48$. The thickness of the composite cathode is $30 \mu \text{m}$. The anode is modeled as a planar lithium metal foil with a thickness of $10 \mu \text{m}$. A thin current collector foil with a thickness of $2 \mu \text{m}$ is attached to both electrodes. The separator is located between the electrodes with a thickness of $20 \mu \text{m}$. It consists of the same solid electrolyte grains as in the composite cathode. We construct a planar grain boundary where two solid electrolyte grains intersect. Thereby, a connected network of grain boundaries is created (see \cref{fig:realistic_geometry_gb}). The average porosity of the porous separator and the porous composite cathode is $\epsilon=0.15$. The lateral edge length is 36 $\mu$m to limit the size of the computational domain.
\subsubsection{Materials}
The active material of the cathode is NMC622, the solid electrolyte is LLTO, the anode is lithium metal, and the current collectors are copper and aluminum, respectively. The material parameters are summarized in \cref{table:material_parameters_realistic}.
\subsubsection{Boundary and initial conditions}
A discharge scenario is simulated. Therefore, the initial concentration in the electrodes is chosen to represent a charged state, i.e $c_{\text{c},0} = 21,000 \ \frac{\text{mol}}{\text{m}^3}$ and $c_{\text{a},0} = \frac{\rho_\text{a}}{M_\text{a}} = 76,900 \ \frac{\text{mol}}{\text{m}^3}$. A constant voltage of $\Phi = 0 \ \text{V}$ is enforced at the current collector on the anode side, and a constant current is applied to the current collector at the cathode side, such that the battery cell is discharged with a c rate of 0.1C until the cut-off voltage between both current collectors of~$\Delta \Phi=2.7 \ \text{V}$ is reached. 
%
\subsubsection{Results}
\begin{figure}[ht]
    \centering
    \includegraphics[width = 0.4\textwidth]{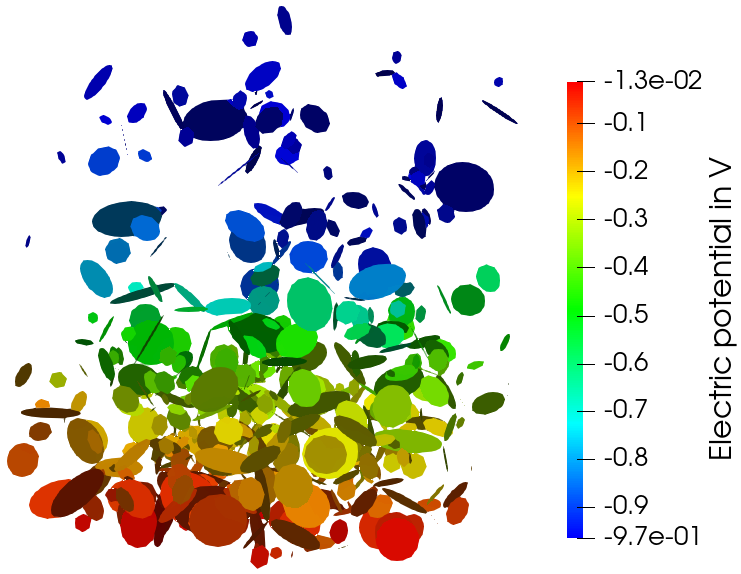}
    \caption{Electric potential in the grain boundaries at the end of discharge.}
    \label{fig:realistic_el_pot}
\end{figure}

\begin{figure}[ht]
    \centering
    \begin{subfigure}[b]{0.45\textwidth}
        \centering
        \input{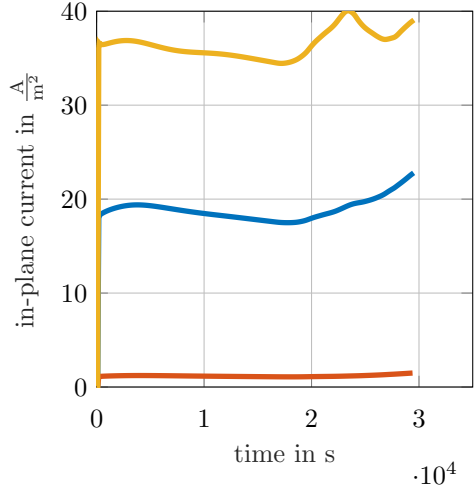}
        \caption{In-plane current for different values of the ionic conductivity in the grain boundary: The standard value ($\kappa_\text{gb} = 1.88 \cdot 10^{-2} \frac{\text{S}}{\text{m}}$, blue curve), a lower value ($\kappa_\text{gb} = 1.88 \cdot 10^{-4} \frac{\text{S}}{\text{m}}$, red curve), and a higher value ($\kappa_\text{gb} = 1.88 \frac{\text{S}}{\text{m}}$, yellow curve).}
        \label{fig:realistic_in_plane_flux}
    \end{subfigure}
    \hfill
    \begin{subfigure}[b]{0.45\textwidth}
        \centering
        \input{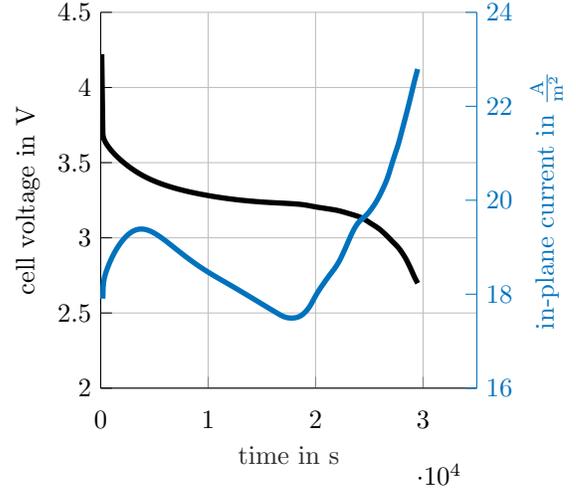}
        \caption{Zoom into the in-plane current for the standard value of the ionic conductivity. Additionally, the cell voltage is shown.}
        \label{fig:realistic_cell_voltage_in_plane_flux_combined}
    \end{subfigure}
    \caption{In-plane current over time.}
\end{figure}

\begin{figure}[ht]
    \centering
    \includegraphics[width = 1.0\textwidth]{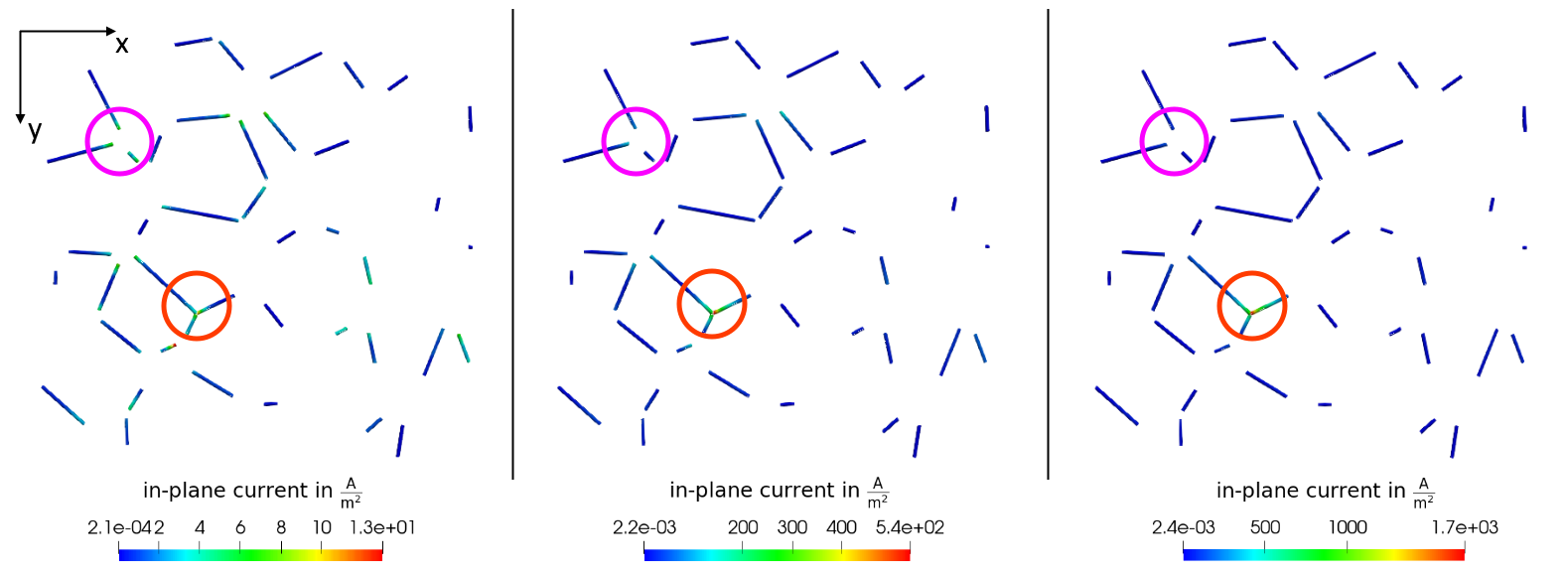}
    \caption{Electric current inside of slices through grain boundaries for different values of the ionic conductivity of the grain boundary (from left to right: small to large, $\kappa_\text{gb} = 1.88 \cdot \left[10^{-4}, 10^{-2}, 1 \right] \frac{\text{S}}{\text{m}}$) at the end of discharge. Note the different limits of the color bar that indicate the different orders of magnitude of the current for the different ionic conductivities. The red circles indicate the intersection between three grain boundaries. The purple circle indicates non-intersecting grain boundaries.}
    \label{fig:realistic_slice}
\end{figure}

The electric potential in the grain boundaries at the end of discharge is shown in \cref{fig:realistic_el_pot}.
A decrease in the electric potential from the anode to the cathode occurs due to the battery cell being discharged. The current along the grain boundary is computed as $\vec{i}_\text{t} = \vec{i} - ({\vec{n}^\text{gb}}^\text{T} \vec{i}) \vec{n}^\text{gb}$ to investigate the influence of the grain boundary in more detail. The averaged in-plane current is $\bar{i}_\text{t} = \frac{\int_{\Gamma_\text{gb}} |\vec{i}_\text{t}| \ \text{d}\Gamma}{\int_{\Gamma_\text{gb}}  \text{d}\Gamma}$. It is shown in \cref{fig:realistic_in_plane_flux} for different values of the ionic conductivity of the grain boundary. 
We observe that the in-plane current increases with higher ionic conductivities as the conduction path inside the grain boundaries becomes more preferred compared to conduction in the grains. Furthermore, it can be seen that the in-plane current does not remain constant but changes over time. This can be attributed to the inhomogeneous lithiation of the cathode, which results in an inhomogeneous electronic conductivity in the cathode (see \cref{eq:nmc_cond}) and an inhomogeneous distribution of the equilibrium potential at the interface between cathode and solid electrolyte and, therefore, to shifted optimal conduction paths during discharging. If, thereby, the impedance in the cathode is increased, conduction in the solid electrolyte and, hence, along the grain boundaries become more favored and vice versa. Thus, the current along the grain boundary changes over time. Additionally, the cell voltage is shown in \cref{fig:realistic_cell_voltage_in_plane_flux_combined} together with a zoom into the in-plane current for the standard value of the ionic conductivity in the grain boundary to highlight the dependence of the state of charge of the in-plane current. \\
In \cref{fig:realistic_slice}, the geometrically resolved magnitude of the in-plane current inside of the grain boundaries is shown for different values of the ionic conductivities at the end of the discharge in a slice close to the anode. 
Note that the dominant direction of the current is out of the shown plane, i.e., in the z-direction of \cref{fig:realistic_geometry_full}. An increase in current is observed for higher conductivities (indicated by the different color bars in \cref{fig:realistic_slice}). Moreover, an increased current is also observed at the intersection between more than two grain boundaries, marked with red circles in \cref{fig:realistic_slice}. This is caused by the current that is merged from two grain boundaries into one grain boundary; therefore, this increase is not observed in non-intersecting grain boundaries (purple circles in \cref{fig:realistic_slice}). This local increase in current could be unfavored as it could initiate the development of dendrites or local plating \cite{Liu2020b,Singh2022a,Cao2020}. Beyond this physical insight, this also highlights the necessity of consistent constraint enforcement at the intersections to properly resolve the increase in current there.
\newpage
\section{Summary}
A modeling approach is presented to incorporate the ionic conduction along grain boundaries into a continuum model for solid-state batteries that geometrically resolves the microstructure. Based on a formulation to represent transport in thin layers of solid-state batteries~\cite{Sinzig2023a}, the grain boundaries are reduced to a two-dimensional manifold. This reduction raises the question of how to guarantee the conservation of mass and charge at locations where more than two grain boundaries, modeled as two-dimensional manifolds, intersect. In terms of ionic conduction inside of the grain boundaries, this means a unique electric potential and a net current of zero. These constraints are enforced by treating the electric potential and the electric current as independent unknowns within the system of equations. This formulation comes with some numerical challenges like the solution of a saddle point system, which we discuss in this work together with solution strategies for them. \\
We show the fulfillment of the formulated constraints as well as the convergence of the numerical error. Afterwards, we discuss extreme cases for the limit of infinite and zero ionic conductivity of the grain boundaries in terms of their influence on the voltage drop within the solid electrolyte. We observe that the novel formulation is able to cover both extreme cases as well as a smooth transition between them. Finally, we show the applicability of the model to realistic microstructures, extract the current along the grain boundaries, and find an increase in the magnitude of the flux at the intersection of grain boundaries, which could be attributed to degradation mechanisms in further studies. Furthermore, the results of impedance measurements could be classified into contributions from grain boundaries and bulk domains. \\
The outlined formulation is exemplarily shown for ionic conduction in the grain boundaries. However, recent publications indicate that also electronic conduction~\cite{Song2019} may occur in grain boundaries of solid-state batteries. With the model established in this work, this and also various other transport phenomena in grain boundaries can be incorporated into the model together with advanced interface kinetics to also model, e.g., lithium deposition in the grain boundaries.
\section*{Funding}
We gratefully acknowledge support by the Bavarian Ministry of Economic Affairs, Regional Development and Energy [project ``Industrialisierbarkeit von Festk\"orperelektrolytzellen''] and the German Federal Ministry of Education and Research [FestBatt~2 (03XP0435B)].
\begin{appendices}
    \section{Discretization of the equations in the grain boundary}
\label{sec:remarks_constr}
For discretizing \cref{eq:residual_std,eq:residual_constr_phi,eq:residual_constr_i} the unknowns~$\Phi_\text{gb}, \vec{i}_\text{gb}$ and the test functions~$w_\Phi, \vec{w}_i$, polynomial shape functions that are organized in the matrix $\mat{N}$, i.e., $\left[\Phi_\text{gb}, \vec{i}_\text{gb}, w_\Phi, \vec{w}_i \right]^\text{T} = \mat{N} \left[\hat{\Phi}_\text{gb}, \vec{\hat{i}}_\text{gb}, \vec{\hat{w}}_\Phi, \vec{\hat{w}}_i \right]^\text{T}$ are employed. In this work, linear Lagrangian polynomials are used to form the space of the shape functions. The spatially discretized forms of the weighted residua are
\begin{subequations}
\begin{align}
    &\hat{R}_\text{gb,std} = \int_{\Gamma_\text{gb}} \nabla_\Gamma (\vec{\hat{w}}_\Phi^\text{T} \vec{N}^\text{T}) \mat{N} \vec{\hat{i}}_\text{gb} \text{d} \Gamma + \int_{\Gamma_\text{gb}} (\vec{\hat{w}}_\Phi^\text{T} \vec{N}^\text{T}) s_{\rho_\text{gb}} \text{d} \Gamma + \int_{\Gamma_\text{gb}} \vec{\hat{w}}_i^\text{T} \mat{N}^\text{T} (\mat{N} \vec{\hat{i}}_\text{gb} + \kappa_\text{gb} \nabla_\Gamma (\vec{N} \vec{\hat{\Phi}}_\text{gb})) \text{d} \Gamma, \\
    &\hat{R}_{\text{gb,constr}_\Phi} = \sum_{j=1}^{n-1} \left[ \int_{\partial \Gamma_\text{gb}} {\vec{\hat{w}}_{{\lambda_\Phi,j}}}^\text{T} \mat{N}^\text{T} \mat{N} \left(\vec{\hat{\Phi}}_\text{m} - \vec{\hat{\Phi}}_{\text{s},j} \right) \text{d} \partial \Gamma + 
    \int_{\partial \Gamma_\text{gb}} \left({\vec{\hat{w}}_\Phi}_\text{m}^\text{T} - {{\vec{\hat{w}}_\Phi}_{\text{s},j}}^\text{T} \right) \mat{N}^\text{T} \mat{N} \vec{\hat{\lambda}}_{\Phi,j} \text{d} \partial \Gamma \right], \\
    &\hat{R}_{\text{gb,constr}_i} = \int_{\partial \Gamma_\text{gb}} {\vec{\hat{w}}_{\lambda}}_i^\text{T} \mat{N}^\text{T} \left( \mat{N} \vec{\hat{i}}_\text{s}^1 + \mat{C}_i \mat{N}\vec{\hat{i}}_\text{m} \right) \text{d} \partial \Gamma + \int_{\partial \Gamma_\text{gb}} \left( {{\vec{\hat{w}}_i}_\text{s}^1}^\text{T} \mat{N}^\text{T} + {\vec{\hat{w}}_i}_\text{m}^\text{T} \mat{N}^\text{T} \mat{C}_i^\text{T} \right) \mat{N} \vec{\hat{\lambda}}_i \text{d} \partial \Gamma.
\end{align}
\end{subequations}
The residual $\hat{R}_\text{gb} = \hat{R}_\text{gb,std} + \hat{R}_{\text{gb,constr}_\Phi} + \hat{R}_{\text{gb,constr}_i}$ is reorganized w.r.t. nodal values of the discretized functions by introducing the subscript~'i' for interior values that are neither assigned to master nor to slave nodes
\begin{equation}
    \begin{split}
    \hat{R}_{\text{gb}_\text{i,m,s}} = {\vec{\hat{w}}_\Phi}_\text{i,m,s}^\text{T} \left( \mat{D} {\vec{\hat{i}}_\text{gb}}_\text{i,m,s} + \vec{b} \right) + {\vec{\hat{w}}_\Phi}_\text{m}^\text{T} \sum_{j=1}^{n-1} \mat{M}_\text{b} \vec{\hat{\lambda}}_{\Phi,j} - \sum_{j=1}^{n-1} {{\vec{\hat{w}}_\Phi}_{\text{s},j}}^\text{T} \mat{M}_\text{b} \vec{\hat{\lambda}}_{\Phi,j}\\
    + {\vec{\hat{w}}_i}_\text{i,m,s}^\text{T} \left( \mat{M} \vec{\hat{i}}_\text{gb} + \mat{G} \vec{\hat{\Phi}}_\text{gb} \right) + {{\vec{\hat{w}}_i}_\text{s}^1}^\text{T} \mat{M}_\text{b} \vec{\hat{\lambda}}_i + {\vec{w}_i}_\text{m}^\text{T} \mat{M}_\text{c}^\text{T} \vec{\hat{\lambda}}_i \\
    + \sum_{j=1}^{n-1} {\vec{\hat{w}}_{\lambda_{\Phi,j}}}^\text{T} \mat{M}_\text{b} \left( \vec{\hat{\Phi}}_\text{m} - \vec{\hat{\Phi}}_\text{s}^j \right) + {\vec{\hat{w}}_\lambda}_i^\text{T} \left( \mat{M}_\text{b} \vec{\hat{i}}_\text{s}^1 + \mat{M}_\text{c} \vec{\hat{i}}_\text{m} \right) = \\
    \vec{w}_{\Phi_\text{i}}^\text{T} {\vec{R}_\text{gb}}_{\Phi_\text{i}} + \vec{w}_{\Phi_\text{m}}^\text{T} {\vec{R}_\text{gb}}_{\Phi_\text{m}} + \vec{w}_{\Phi_\text{s}}^\text{T} {\vec{R}_\text{gb}}_{\Phi_\text{s}} + \vec{w}_{i_\text{i}}^\text{T} {\vec{R}_\text{gb}}_{i_\text{i}} \\ + \vec{w}_{i_\text{m}}^\text{T} {\vec{R}_\text{gb}}_{i_\text{m}} + \vec{w}_{i_\text{s}}^\text{T} {\vec{R}_\text{gb}}_{i_\text{s}} + \vec{w}_{\lambda_\Phi}^\text{T} {\vec{R}_\text{gb}}_{\lambda_\Phi} + \vec{w}_{\lambda_i}^\text{T} {\vec{R}_\text{gb}}_{\lambda_i} = 0,
    \end{split}
\end{equation}
with $\mat{D} = \int_{\Gamma_\text{gb}} \nabla_\Gamma \vec{N}^\text{T} \mat{N} \text{d} \Gamma$, $\vec{b} = \int_{\Gamma_\text{gb}} \vec{N}^\text{T} s_{\rho_\text{gb}} \text{d} \Gamma$, $\mat{M}_\text{b} = \int_{\partial \Gamma_\text{gb}} \mat{N}^\text{T} \mat{N} \text{d} \partial \Gamma$, $\mat{M} = \int_{\Gamma_\text{gb}} \mat{N}^\text{T} \mat{N} \text{d} \partial \Gamma$, $\mat{M}_\text{c} = \int_{\partial \Gamma_\text{gb}} \mat{N}^\text{T} \mat{C}_i \mat{N} \text{d} \partial \Gamma$, and $\mat{G} = \int_{\Gamma_\text{gb}} \kappa_\text{gb} \mat{N}^\text{T} \nabla_\Gamma \vec{N} \text{d} \Gamma$.
The test functions~$\vec{\hat{w}}$ have arbitrary values, s.t. each term has to be individually zero, i.e.
\begin{subequations}
\begin{equation}
    \vec{\bar{R}}_\text{gb} = 
    \begin{bmatrix}
        {{\vec{R}_\text{gb}}_\Phi}_\text{i} &
        {{\vec{R}_\text{gb}}_\Phi}_\text{m} &
        {{\vec{R}_\text{gb}}_\Phi}_\text{s} &
        {{\vec{R}_\text{gb}}_i}_\text{i} &
        {{\vec{R}_\text{gb}}_i}_\text{m} &
        {{\vec{R}_\text{gb}}_i}_\text{s} &
        {{\vec{R}_\text{gb}}_\lambda}_\Phi &
        {{\vec{R}_\text{gb}}_\lambda}_i &
    \end{bmatrix}^\text{T}
    = \vec{0}.
\end{equation}
This nonlinear, algebraic system of equations is iteratively solved for the unknowns organized in a vector~$\vec{\bar{\Psi}}_\text{gb} = [\vec{\hat{\Phi}}_\text{i}, \vec{\hat{\Phi}}_\text{m}, \vec{\hat{\Phi}}_\text{s}, \vec{\hat{i}}_\text{i}, \vec{\hat{i}}_\text{m}, \vec{\hat{i}}_\text{s}, \vec{\hat{\lambda}}_\Phi, \vec{\hat{\lambda}}_i]^\text{T}$ by the Newton-Raphson scheme~$\left.\parder{\vec{\bar{R}}_\text{gb}}{\vec{\bar{\Psi}}_\text{gb}}\right|_i \Delta {\vec{\bar{\Psi}}_\text{gb}}_i = - \vec{\bar{R}}_\text{gb}({\vec{\bar{\Psi}}_\text{gb}}_i)$, ${\vec{\bar{\Psi}}_\text{gb}}_{i+1} = {\vec{\bar{\Psi}}_\text{gb}}_i + \Delta {\vec{\bar{\Psi}}_\text{gb}}_i$ where the matrix~$\mat{\bar{K}}_\text{gb} = \left.\parder{\vec{\bar{R}}_\text{gb}}{{\vec{\bar{\Psi}}_\text{gb}}}\right|_i$ is given by
\begin{equation}
    \mat{K}_\text{gb} = 
    \begin{bmatrix}
        \parder{\vec{b}}{\Phi_\text{i}} & \mat{0} & \mat{0} & \mat{D}_\text{ii} & \mat{D}_\text{im} & \mat{D}_\text{is} & \mat{0} & \mat{0}\\
        \mat{0} & \parder{\vec{b}}{\Phi_\text{m}} & \mat{0} & \mat{D}_\text{mi} & \mat{D}_\text{mm} & \mat{0} & \sum_{j=1}^{n-1} \mat{M}_\text{b} & \mat{0}\\
        \mat{0} & \mat{0} & \parder{\vec{b}}{\Phi_\text{s}} & \mat{D}_\text{si} & \mat{0} & \mat{D}_\text{ss} & \sum_{j=1}^{n-1} -\mat{M}_\text{b} & \mat{0}\\
        \mat{G}_\text{ii} & \mat{G}_\text{im} & \mat{G}_\text{is} & \mat{M}_\text{ii} & \mat{M}_\text{im} & \mat{M}_\text{is} & \mat{0} & \mat{0} \\
        \mat{G}_\text{mi} & \mat{G}_\text{mm} & \mat{0} & \mat{M}_\text{mi} & \mat{M}_\text{mm} & \mat{0} & \mat{0} & \mat{M}_\text{c}^\text{T} \\
        \mat{G}_\text{si} & \mat{0} & \mat{G}_\text{ss} & \mat{M}_\text{si} & \mat{0} & \mat{M}_\text{ss} & \mat{0} & \mat{M}_\text{b} \\
        \mat{0} & \sum_{j=1}^{n-1} \mat{M}_\text{b} & \sum_{j=1}^{n-1} - \mat{M}_\text{b} & \mat{0} & \mat{0} & \mat{0} & \mat{0} & \mat{0} \\
        \mat{0} & \mat{0} & \mat{0} & \mat{0} & \mat{M}_\text{c} & \mat{M}_\text{b} & \mat{0} & \mat{0}
    \end{bmatrix}
    .
\end{equation}
\end{subequations}
The subscripts denote the interior, master, or slave side matrices. The Lagrange multipliers, as well as the slave side values, are removed from the global system of equations by static condensation, and the final system of linear equations is
\begin{subequations}
\begin{equation}
    \begin{split}
    &\mat{K}_\text{gb} = \\
    &\begin{bmatrix}
        \parder{\vec{b}}{\Phi_\text{i}} & \mat{0} & \mat{D}_\text{ii} & \mat{D}_\text{im} - \mat{D}_\text{is} \mat{C}_i\\
        \mat{0} & \parder{\vec{b}}{\Phi_\text{m}} + \parder{\vec{b}}{\Phi_\text{s}} &\mat{D}_\text{mi} + \mat{D}_\text{si} & \mat{D}_\text{mm} - \mat{D}_\text{ss} \mat{C}_i\\
        \mat{G}_\text{ii} & \mat{G}_\text{im} + \mat{G}_\text{is} & \mat{M}_\text{ii} & \mat{M}_\text{im} -  \mat{M}_\text{is} \mat{C}_i \\
        \mat{G}_\text{mi} - \mat{C}_i^\text{T} \mat{G}_\text{si} & \mat{G}_\text{mm} - \mat{C}_i^\text{T} \mat{G}_\text{ss}  & \mat{M}_\text{mi} - \mat{C}_i^\text{T} \mat{M}_\text{mi} & \mat{M}_\text{mm} + \mat{C}_i^\text{T} \mat{M}_\text{ss} \mat{C}_i
    \end{bmatrix}
    =
    \begin{bmatrix}
        {\mat{K}_\text{gb}}_{\Phi \Phi} & {\mat{K}_\text{gb}}_{\Phi i} \\
        {\mat{K}_\text{gb}}_{i \Phi} & {\mat{K}_\text{gb}}_{i i}
    \end{bmatrix},
\end{split}
\end{equation}
\begin{equation}
    \vec{R}_\text{gb} =
    \begin{bmatrix}
        {{\vec{R}_\text{gb}}_\Phi}_\text{i} &
        {{\vec{R}_\text{gb}}_\Phi}_\text{m} + {{\vec{R}_\text{gb}}_\Phi}_\text{s} &
        {{\vec{R}_\text{gb}}_i}_\text{i} &
        {{\vec{R}_\text{gb}}_i}_\text{m} - \mat{C}_\text{s}^\text{T} {{\vec{R}_\text{gb}}_i}_\text{s}
    \end{bmatrix}^\text{T} =
    \begin{bmatrix}
        {\vec{R}_\text{gb}}_\Phi &
        {\vec{R}_\text{gb}}_i
    \end{bmatrix}^\text{T}
    ,
\end{equation}
\end{subequations}
with the condensed vector of unknowns $\vec{\Psi}_\text{gb} = [\vec{\hat{\Phi}}_\text{i}, \vec{\hat{\Phi}}_\text{m}, \vec{\hat{i}}_\text{i}, \vec{\hat{i}}_\text{m}]^\text{T}$.
\section{Discretization of the equations in the bulk domains}
\label{sec:remarks_bulk}
For discretizing \cref{eq:residual_bulk}, the unknowns~$\Phi_\text{bulk}, c$ and the test functions~$w_\Phi$ and $w_c$ are as well discretized with polynomial shape functions $\mat{N}$, i.e., $\left[\Phi_\text{bulk}, c, w_\Phi, w_c \right]^\text{T} = \mat{N} \left[\hat{\Phi}_\text{bulk}, \vec{\hat{c}}, \vec{\hat{w}}_\Phi, \vec{\hat{w}}_c \right]^\text{T}$. Again, linear Lagrangian polynomials are used to form the space of the shape functions. 
\begin{equation}
    \begin{split}
    R_\text{bulk} = 
    \vec{w}_\Phi^\text{T} \kappa_\text{el} \int_{\Omega_\text{el}} \nabla \mat{N}^\text{T} \nabla \mat{N} \text{d} \Omega \vec{\hat{\Phi}} + \vec{w}_\Phi^\text{T} \int_{{\Gamma_\text{el}}_\text{h}} \mat{N}^\text{T} \bar{i} \text{d} \Gamma \\
    + \vec{w}_\Phi^\text{T} \sigma \int_{\Omega_\text{ed,cc}} \nabla \mat{N}^\text{T} \nabla \mat{N} \text{d} \Omega \vec{\hat{\Phi}} + \vec{w}_\Phi^\text{T} \int_{{\Gamma_\text{ed,cc}}_\text{h}} \mat{N}^\text{T} \bar{i} \text{d} \Gamma \\
    +\vec{w}_c^\text{T} \int_{\Omega_\text{ed}} \mat{N}^\text{T} \mat{N} \text{d} \Omega \parder{\vec{\hat{c}}}{t} + \vec{w}_c^\text{T} D \int_{\Omega_\text{ed}} \nabla \mat{N}^\text{T} \nabla \mat{N} \text{d} \Omega \vec{\hat{c}} + \vec{w}_c^\text{T} \int_{{\Gamma_\text{ed}}_\text{h}} \mat{N}^\text{T} \bar{j} \text{d} \Gamma = 0.
    \end{split}
\end{equation}
The linearized system of equations for the Newton-Raphson scheme in the bulk domains is given as
\begin{subequations}
\begin{align}
    \vec{R}_\text{bulk} =
    \begin{bmatrix}
         \{\kappa,\sigma\} \mat{K} \vec{\hat{\Phi}} + \int_\Gamma \mat{N}^\text{T} \bar{i} \text{d} \Gamma \\
         \mat{S} \parder{\vec{\hat{c}}}{t} + D \mat{K} \vec{\hat{c}} + \int_\Gamma \mat{N}^\text{T} \bar{j} \text{d} \Gamma
    \end{bmatrix} = 
    \begin{bmatrix}
        {\vec{R}_\text{bulk}}_\Phi \\
        {\vec{R}_\text{bulk}}_c
    \end{bmatrix}, \\
    \mat{K}_\text{bulk} =
    \begin{bmatrix}
        \{\kappa,\sigma\} \mat{K} + \int_\Gamma \mat{N}^\text{T} \parder{\bar{i}}{\vec{\hat{\Phi}}} \text{d} \Gamma & \int_\Gamma \mat{N}^\text{T} \parder{\bar{i}}{\vec{\hat{c}}} \text{d} \Gamma \\
        \int_\Gamma \mat{N}^\text{T} \parder{\bar{j}}{\vec{\hat{\Phi}}} \text{d} \Gamma & \mat{S} \parder{\left(\parder{\vec{\hat{c}}}{t}\right)}{\vec{\hat{c}}} + D \mat{K} + \int_\Gamma \mat{N}^\text{T} \parder{\bar{j}}{\vec{\hat{c}}} \text{d} \Gamma
    \end{bmatrix} =
    \begin{bmatrix}
        {\mat{K}_\text{bulk}}_{\Phi \Phi} & {\mat{K}_\text{bulk}}_{\Phi c} \\
        {\mat{K}_\text{bulk}}_{c \Phi} & {\mat{K}_\text{bulk}}_{c c}
    \end{bmatrix},
\end{align}
\end{subequations}
with $\mat{K} = \int_\Omega \nabla \mat{N}^\text{T} \nabla \mat{N} \text{d} \Omega$, $\mat{S} = \int_\Omega \mat{N}^\text{T} \mat{N} \text{d} \Omega$ and a time discretization scheme for $\parder{\vec{\hat{c}}}{t}$; in this case the one-step theta method $\parder{\vec{\hat{c}}}{t} \approx \frac{\vec{\hat{c}}_{t+1}-\vec{\hat{c}}_t}{\Delta t} = \Theta \left(-D \mat{K} \vec{\hat{c}}_{t+1} - \int_\Gamma \mat{N}^\text{T} \parder{\bar{j}}{\vec{\hat{c}}} \text{d} \Gamma |_{t+1} \right) + (1-\Theta) \left(-D \mat{K} \vec{\hat{c}}_t - \int_\Gamma \mat{N}^\text{T} \parder{\bar{j}}{\vec{\hat{c}}} \text{d} \Gamma|_t \right)$. The expression $\{\kappa,\sigma\}$ denotes the ionic or electronic conductivity in the respective domains.
\section{Block Gauss-Seidel algorithm for saddle-point systems}
The Block Gauss-Seidel algorithm (\cref{alg:BGS}) is used to iteratively solve the linear system of equations $\mat{K} \Delta \vec{x} = \vec{R}$ until $\text{norm} \left(\mat{E}_\text{r} \mat{K} \mat{E}_\text{c} \Delta \vec{x} - \mat{E}_\text{r} \vec{R} \right) < \epsilon$, with a tolerance~$\epsilon$ Thereby, only systems of equations that contain the $n$ blocks on the main diagonal need to be solved. If this sub-system contains a saddle-point structure, a direct solver is employed. The full system of equations is equilibrated by row~($\mat{E}_\text{r}$) and column~($\mat{E}_\text{c}$) multiplication to improve its condition. We chose both matrices as diagonal matrices containing the reciprocal of the largest value of the respective row or column within each sub-block.
\begin{algorithm}
    \caption{Block Gauss-Seidel with saddle-point structure}
    \begin{algorithmic}        
        \State $\mat{A} \gets \mat{E}_\text{r} \mat{K} \mat{E}_\text{c}$ \Comment{Equilibration}
        \State $\vec{c} \gets \mat{E}_\text{r} \vec{R}$
        \State $\vec{y}_i \gets \vec{0} \quad \forall \ i \in n$
        \State $h \gets 0$
        \While{$r > \epsilon$}
        \State $h \gets h + 1$
            \For{$i \in \{1,\dots,n\}$} \Comment{Loop over main-diagonal blocks}
            \State $\vec{g} \gets \vec{c}_i$
                \For{$j \in \{1,\dots,n\}$} \Comment{Loop over off-diagonal blocks}
                    \If{$j < i$}
                        \State $\vec{g} \gets \vec{g} - \mat{A}_{ij} \vec{y}_j^{h-1}$
                    \ElsIf{$j > i$}
                        \State $\vec{g} \gets \vec{g} - \mat{A}_{ij} \vec{y}_j^h$
                    \EndIf
                \EndFor
            \State $\vec{y}_i \gets$ \Call{Solve}{$\mat{A}_{ii}, \vec{g}, i$}
            \EndFor
        \State $r \gets \text{norm}(\mat{A}\vec{y}-\vec{c})$
        \EndWhile
        \State $\Delta \vec{x} \gets \mat{E}_\text{c} \vec{y}$ \Comment{Substitution from equilibration}
        \State
        \Function{Solve}{$\mat{A}, \vec{g}$, $i$}
        \If{$i == \Call{Block}{\text{gb}}$}
            \State \Return $\Call{DirectSolver}{\mat{A}, \vec{g}}$
        \Else
            \State \Return $\Call{AMGSolver}{\mat{A}, \vec{g}}$
        \EndIf
        \EndFunction
    \end{algorithmic}
    \label{alg:BGS}
\end{algorithm}
\section{Material parameters}
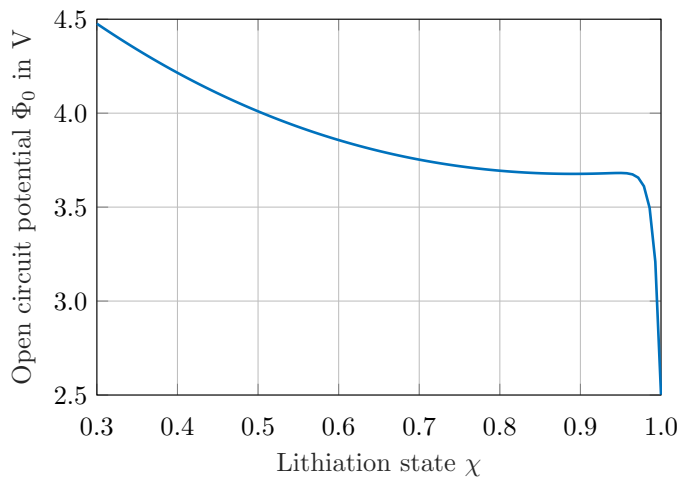
\begin{figure}[H]
    \centering
    \definecolor{mycolor1}{rgb}{0.00000,0.44700,0.74100}%

\begin{tikzpicture}

\begin{axis}[%
width=7.5cm,
height=5cm,
scale only axis,
xmin=0.3,
xmax=1.0,
x tick label style={
/pgf/number format/.cd,
fixed,
fixed zerofill,
precision=1,
/tikz/.cd,
yshift=-.5em},
xlabel style={font=\color{white!15!black}},
xlabel={Lithiation state $\chi$},
ymin=2.5,
ymax=4.5,
ytick={2.5,3.0,3.5,4.0,4.5},
scaled y ticks = false,
y tick label style={
/pgf/number format/.cd,
fixed,
fixed zerofill,
precision=1,
/tikz/.cd},
ylabel style={font=\color{white!15!black}},
ylabel={Open circuit potential $\Phi_0$ in V},
axis background/.style={fill=white},
xmajorgrids,
ymajorgrids,
legend style={at={(0.97,0.03)}, anchor=south east, legend cell align=left, align=left, draw=white!15!black}
]
\addplot [color=mycolor1, line width=1.0pt]
  table[row sep=crcr]{%
0.3 4.47619927240105\\
0.307070707070707 4.45582451376056\\
0.314141414141414 4.43574107056826\\
0.321212121212121 4.41594886951643\\
0.328282828282828 4.39644765830068\\
0.335353535353535 4.37723702746023\\
0.342424242424242 4.35831642956736\\
0.349494949494949 4.33968519611955\\
0.356565656565657 4.32134255243552\\
0.363636363636364 4.30328763081285\\
0.370707070707071 4.28551948216771\\
0.377777777777778 4.26803708634665\\
0.384848484848485 4.25083936127388\\
0.391919191919192 4.23392517107541\\
0.398989898989899 4.21729333330258\\
0.406060606060606 4.20094262536124\\
0.413131313131313 4.18487179023908\\
0.42020202020202 4.16907954161204\\
0.427272727272727 4.15356456840006\\
0.434343434343434 4.13832553883432\\
0.441414141414141 4.12336110408984\\
0.448484848484848 4.10866990153148\\
0.455555555555555 4.09425055761504\\
0.462626262626263 4.08010169048072\\
0.46969696969697 4.06622191227157\\
0.476767676767677 4.05260983120601\\
0.483838383838384 4.03926405342996\\
0.490909090909091 4.02618318467165\\
0.497979797979798 4.0133658317191\\
0.505050505050505 4.00081060373859\\
0.512121212121212 3.98851611345006\\
0.519191919191919 3.97648097817383\\
0.526262626262626 3.96470382076164\\
0.533333333333333 3.95318327042322\\
0.54040404040404 3.94191796345897\\
0.547474747474747 3.93090654390784\\
0.554545454545454 3.92014766411865\\
0.561616161616162 3.90963998525251\\
0.568686868686869 3.89938217772274\\
0.575757575757576 3.88937292157861\\
0.582828282828283 3.87961090683814\\
0.58989898989899 3.87009483377496\\
0.596969696969697 3.86082341316362\\
0.604040404040404 3.85179536648736\\
0.611111111111111 3.84300942611191\\
0.618181818181818 3.8344643354287\\
0.625252525252525 3.82615884897028\\
0.632323232323232 3.81809173250078\\
0.639393939393939 3.81026176308373\\
0.646464646464646 3.80266772912953\\
0.653535353535354 3.79530843042446\\
0.660606060606061 3.78818267814323\\
0.667676767676768 3.78128929484642\\
0.674747474747475 3.77462711446467\\
0.681818181818182 3.76819498227066\\
0.688888888888889 3.76199175484031\\
0.695959595959596 3.75601630000432\\
0.703030303030303 3.75026749679087\\
0.71010101010101 3.74474423536075\\
0.717171717171717 3.73944541693546\\
0.724242424242424 3.7343699537192\\
0.731313131313131 3.72951676881543\\
0.738383838383838 3.72488479613869\\
0.745454545454545 3.72047298032212\\
0.752525252525252 3.71628027662136\\
0.75959595959596 3.71230565081522\\
0.766666666666667 3.70854807910363\\
0.773737373737374 3.70500654800316\\
0.780808080808081 3.70168005424053\\
0.787878787878788 3.6985676046444\\
0.794949494949495 3.69566821603546\\
0.802020202020202 3.69298091511488\\
0.809090909090909 3.69050473835049\\
0.816161616161616 3.68823873185844\\
0.823232323232323 3.68618195127564\\
0.83030303030303 3.6843334616095\\
0.837373737373737 3.68269233703302\\
0.844444444444444 3.68125766054535\\
0.851515151515151 3.68002852330237\\
0.858585858585859 3.67900402313585\\
0.865656565656566 3.6781832610791\\
0.872727272727273 3.67756533299432\\
0.87979797979798 3.67714930916538\\
0.886868686868687 3.67693418432381\\
0.893939393939394 3.67691875503528\\
0.901010101010101 3.6771013186283\\
0.908080808080808 3.67747893369498\\
0.915151515151515 3.67804560348254\\
0.922222222222222 3.67878781309469\\
0.929292929292929 3.67967356566211\\
0.936363636363636 3.68062544709972\\
0.943434343434343 3.68145445308287\\
0.95050505050505 3.68169741857838\\
0.957575757575758 3.68021762288952\\
0.964646464646465 3.67422357599325\\
0.971717171717172 3.65685842139712\\
0.978787878787879 3.61127770022431\\
0.985858585858586 3.49609989452052\\
0.992929292929293 3.20966204045106\\
1 2.50220468352966\\
};

\end{axis}
\end{tikzpicture}%
    \caption{Open circuit potential of NMC622 as a function of the lithiation state based on~\cite{Kremer2019}.}
    \label{fig:OCP_NMC622}
\end{figure}

\begin{table}[H]
    \renewcommand{\arraystretch}{1.1}
    \centering
    \caption{Material parameters for the simulation with the realistic geometry.}
    \begin{tabular}{c | c | c | c | c }
      \hline
      \textbf{domain}                                            &      \textbf{quantity}   & \textbf{symbol}                        & \textbf{value}                                    & \textbf{source} \\
      \hline
      \multirow{6}{*}{cathode $\Omega_\text{c}$}                & electronic conductivity  & $\sigma$                               & \cref{eq:nmc_cond}                                & \cite{Neumann2020} \\
                                                                 & diffusion coefficient    & $D$                                    & \cref{eq:nmc_diff}                                & \cite{Neumann2020} \\
                                                                 & open circuit potential   & $\Phi_0$                               & \cref{fig:OCP_NMC622}                             & \cite{Kremer2019} \\

                                                                 & max. concentration       & $c_\text{max}$                         & $5.19 \cdot 10^4 \ \frac{\text{mol}}{\text{m}^3}$ & \cite{Neumann2020} \\
                                                                 & max. lithiation          & $\chi_\text{max}$                      & $1$                                               & \cite{Neumann2020} \\
                                                                 & lithiation range         & $[\chi_\text{0\%}, \chi_\text{100\%}]$ & $[1, 0.404]$                                      & defined \\
      \hline
      \multirow{2}{*}{solid electrolyte $\Omega_\text{el}$}     & ionic conductivity       & $\kappa$                               & $7.86 \cdot 10^{-2} \ \frac{\text{S}}{\text{m}}$   & \cite{Braun2017} \\
                                                                 & transference number      & $t_+$                                    & 1                                                 & defined \\
      \hline
      \multirow{2}{*}{grain boundary $\Omega_\text{gb}$}     & ionic conductivity       & $\kappa$                               & $1.88 \cdot 10^{-2} \ \frac{\text{S}}{\text{m}}$   & \cite{Braun2017} \\
                                                                 & transference number      & $t_+$                                    & 1                                                 & defined \\
      \hline
      \multirow{1}{*}{anode $\Omega_\text{a}$}                  & electronic conductivity  & $\sigma$                               & $10^5 \ \frac{\text{S}}{\text{m}}$                & \cite{Neumann2020} \\
      \hline
      \shortstack{current collector \\ anode $\Omega_\text{ac}$}  & electronic conductivity  & $\sigma$                               & $5.81 \cdot 10^7 \ \frac{\text{S}}{\text{m}}$     & \cite{JensFreudenberger2020} \\
      \hline
      \shortstack{current collector \\ cathode $\Omega_\text{cc}$} & electronic conductivity  & $\sigma$                               & $3.77 \cdot 10^7 \ \frac{\text{S}}{\text{m}}$     & \cite{JensFreudenberger2020} \\
      \hline
      \shortstack{interface current \\ coll. - electrode $\Gamma_\text{cs-ed}$}    & interface resistance     & $r_\text{i}$                                    & $2 \cdot 10^{-3} \ \Omega \text{m}^2$                       & defined \\
      \hline
      \shortstack{interface cathode - \\ solid electrolyte $\Gamma_\text{c-el}$}                & exchange current density & $i_0$                                  & $4.98 \ \frac{\text{A}}{\text{m}^2}$              & \shortstack{adapted for \\ NMC622 - $\beta$-LPS \\ from \cite{Schmidt2023} and \cite{Neumann2020}} \\
      \hline
      \shortstack{interface anode - \\ solid electrolyte $\Gamma_\text{an-el}$}        & exchange current density & $i_0$                                  & $8.87 \ \frac{\text{A}}{\text{m}^2}$              & \cite{Neumann2020} \\
      \hline
      \shortstack{interface grain boundary \\ - solid electrolyte $\Gamma_\text{gb-el}$}      & interface resistance     & $r_\text{i}$                                    & $2.0 \cdot 10^{-2} \ \Omega \text{m}^2$   & defined  \\
      \hline
    \end{tabular}
    \label{table:material_parameters_realistic}
\end{table}
The electronic conductivity
\begin{equation}
    \sigma(x) = 100 \frac{\text{S}}{\text{m}} \, \exp(-202.90 \, x^4 + 322.38 \, x^3 - 178.23 \, x^2 + 50.06 \, x - 13.47), \label{eq:nmc_cond}
\end{equation}
with $x = 1 - \chi$ and $\chi = \frac{c}{c_\text{max}} \chi_\text{max} \text{det}(\mat{F})$, and the diffusion coefficient of NMC622 are a function of the lithiation state~\cite{Neumann2020}
\begin{equation}
    \begin{split}
        D(\chi) &= \frac{1}{1000} \frac{\text{m}^2}{\text{s}} \exp \Big(9.3764575854 \cdot 10^5 \cdot \chi^9 - 5.4262087319 \cdot 10^6 \cdot \chi^8 \\
        &+ 1.3688556703 \cdot 10^7 \cdot \chi^7 - 1.9734363260 \cdot 10^7 \cdot \chi^6 + 1.7897244160 \cdot 10^7 \cdot \chi^5 \\
        & - 1.0576735297 \cdot 10^7 \cdot \chi^4 + 4.0688465295 \cdot 10^6 \cdot \chi^3 - 9.8167452940 \cdot 10^5 \cdot \chi^2 \\
        &+ 1.3468923578 \cdot 10^5 \cdot \chi - 8.0270847914 \cdot 10^3 \Big) \, .
    \end{split} \label{eq:nmc_diff}
\end{equation}
The open circuit potential~\cite{Kremer2019} of NMC622 is shown in~\cref{fig:OCP_NMC622}.
 \noindent

\end{appendices}
%
\bibliographystyle{IEEEtran}
\bibliography{literature}
\end{document}